\documentstyle[preprint,prc,eqsecnum,aps,epsf]{revtex}

\tightenlines   

\newcommand{\ffat}[1]{\mbox {\boldmath $#1$}}

\begin{document}
\draft

\title{Nucleon-Nucleon Scattering in a Three Dimensional
Approach}

\author{ I.~Fachruddin$^{\star}$, Ch.~Elster$^{\dagger}$,
W.~Gl\"ockle$^{\star}$} 
\address{ $^{\star}$Institute for Theoretical Physics II, 
Ruhr-University Bochum, D-44780 Bochum, Germany.}
\address{$^{\dagger}$ Institut f\"ur Kernphysik, Forschungszentrum
J\"ulich, D-52425 J\"ulich, and
Institute of Nuclear and
Particle Physics, Ohio University,
Athens, OH 45701}

\vspace{10mm}

\date{\today}

\maketitle

\begin{abstract}
The nucleon-nucleon (NN) t-matrix is calculated directly as function of two 
vector momenta
for different realistic NN potentials. To facilitate this a
 formalism is developed for solving the 
two-nucleon Lippmann-Schwinger equation in momentum space without employing a 
partial wave
decomposition. The total  spin is treated in a helicity representation. 
Two different 
realistic NN interactions, one defined in momentum space and one in
coordinate space,  are presented in a form suited for this formulation.
The angular and momentum dependence of the full amplitude is studied and displayed.
A partial wave decomposition of the full amplitude it carried out to compare the
presented results with the well known phase shifts provided by those interactions.

\end{abstract}
\vspace{10mm}


\pagebreak

 
\narrowtext
 

\section{Introduction}

Experience in three- and four-nucleon calculations \cite{1,2,3}
shows that the standard treatment
based on a partial wave projected momentum space basis is quite successful but
also rather tedious, since each building block requires extended
algebra. For example, 
the representation of the various permutation operators (i.e. 
the transformations among angular momenta belonging to different
Jacobi momenta) is very involved \cite{4} and requires
intricate numerical realizations \cite{5}. 
On the other hand, we demonstrated in \cite{trit3d} the relative ease with which
a 3-boson bound state could be calculated in the Faddeev scheme avoiding an
angular momentum decomposition altogether. Instead of solving a
large set of coupled two-dimensional integral equations in 
the standard partial wave framework, 
in a three dimensional (3D) formulation only one single integral
equation in three dimensions had to be solved.
We had the same positive experience with the scattering of three bosons
\cite{scatter1}, 
where the algebraic formulation and numerical implementation
of the one Faddeev-equation was much simpler than the machinery of the
approach based on partial wave decomposition.

The input to these three-body calculations without angular
momentum decomposition are two-body t-matrices, which are off the energy shell.
In addition to the off-shell energy they depend on the magnitudes 
of the initial and final momenta and the angle between the two momenta.
In \cite{9} we showed that the two-body t-matrices can 
be obtained  very easily by solving a two-dimensional Lippmann-Schwinger
(L.S.) equation instead
of preparing and handling quite a few one-dimensional L.S. equations for each
angular momentum state separately. The off-shell energies required in a
three-body calculation lie between the total
three-body energy and minus infinity. 
Interestingly we found in \cite{9} that the
t-matrix at large negative energies is nearly equal to the real part of the
t-matrix at corresponding positive energy. 
This insight, which was also pointed out in \cite{10},
explains directly that for bound state calculations
a surprisingly large number of partial waves is necessary for
convergence.
In addition, recent three-nucleon (3N) Faddeev calculations \cite{11} based on
realistic nucleon-nucleon (NN) forces and performed between 
100-200 MeV nucleon laboratory energy
reach the very limits of present day computational resources, since the number of
angular momentum states involved increases dramatically. 
Regarding all the above considerations it is 
a challenge to incorporate spin into the previous calculations
\cite{trit3d,scatter1,9} and
perform few-nucleon
calculations without angular momentum decomposition. As an aside, 
in the Monte Carlo calculations
in configuration space \cite{12} this is anyhow a standard procedure.

In the case of  NN scattering
calculations without partial wave decomposition
have already been realized several times \cite{13,14}. In those studies
the spin states were treated in their individual m-representations. 
A different approach is based on a helicity representation \cite{15},
where the helicity related to the total two-nucleon spin was introduced.
In the present work
we follow this basic philosophy, however end up with somewhat 
different final equations to be solved.
We apply our formulation to the realistic NN potentials Bonn-B \cite{16},
 as well as the Argonne AV18 \cite{av18} potential
and check the accuracy of our calculations by comparing to
phase shift parameters obtained from corresponding partial wave
calculations.
The potentials we chose for implementation 
are given in an operator form, and thus directly applicable to our
formulation without partial wave decomposition. 
Other modern NN forces, like Nijm I and II \cite{19} or CD-Bonn
\cite{20}, are
parameterized for each angular momentum state separately and thus 
are not as useful for an approach without partial wave decomposition. 

This article is structured in the following way:
In Section~II we present the formalism. 
The helicity representation of the potentials
and the final form of the Lippmann Schwinger equations are displayed in 
Section~III. 
The implementation of two modern realistic NN forces into our formulation is
given in Section~IV. The
connection to the standard  partial wave representation is given in
Section~V.
Our calculations and results are presented in Section VI, and we
conclude in Section~VII.

\section{Formulation for two nucleon scattering based on Helicities}

First, we  introduce a helicity basis for the total spin S of two nucleons.
For the z-axis being quantization axis, the total spin state has the
well known form
\begin{equation}
\label{1}
\left| \hat{z}s\Lambda \right\rangle =\sum _{m_{1}m_{2}}C(\frac{1}{2}\frac{1}{2}S;m_{1}m_{2}\Lambda )\left| \hat{z}\frac{1}{2}m_{1}\right\rangle \left| \hat{z}\frac{1}{2}m_{2}\right\rangle 
\end{equation}
Applying the rotation operator 
\begin{equation}
\label{2}
R(\hat{q})=e^{-iS_{z}\phi }e^{-iS_{y}\theta },
\end{equation}
 where \( S_{z},\, S_{y} \) are components of the spin operator 
\( {\bf S}=\frac{1}{2}({\mbox {\boldmath $\sigma$}}_1+{\mbox {\boldmath$\sigma$}}_2) \),
leads to the general state
\begin{equation}
\label{3}
\left| \hat{q}S\Lambda \right\rangle =R(\hat{q})\left| \hat{z}S\Lambda \right\rangle 
\end{equation}
This is eigenstate to the helicity operator \( {\bf S}\cdot \hat{q} \):
\begin{equation}
\label{4}
{\bf S}\cdot \hat{q}\left| \hat{q}S\Lambda \right\rangle =\Lambda \left| \hat{q}S\Lambda \right\rangle 
\end{equation}
 This follows simply from 
\( R(\hat{q}){\bf S}\cdot \hat{z}R^{-1}(\hat{q})={\bf S}\cdot \hat{q} \).
Next we define momentum-helicity states as 
\begin{equation}
\label{5}
\left| {\bf q};\hat{q}S\Lambda \right\rangle \equiv \left|
{\bf q}\right\rangle \left| \hat{q}S\Lambda \right\rangle ,
\end{equation}
 where \( {\bf q} \) is the relative momentum of the two nucleons.

\noindent
The parity operator P acts as 
\begin{equation}
\label{6}
P\left| {\bf q};\hat{q}S\Lambda \right\rangle =\left| -{\bf q};\hat{q}S\Lambda \right\rangle 
\end{equation}
on the momentum helicity eigenstates.
Consequently parity eigenstates are given as
\begin{equation}
\label{7}
\left| {\bf q};\hat{q}S\Lambda \right\rangle _{\pi }=
\frac{1}{\sqrt{2}}(1+\eta _{\pi }P)\left| 
{\bf q};\hat{q}S\Lambda \right\rangle ,
\end{equation}
 where \( \eta _{\pi }=\pm 1 \).

\noindent
Combining Eq.~(\ref{7}) with two-body isospin states 
\( \left| tm_{t}\right\rangle  \),we introduce
antisymmetrized two-nucleon states as
\begin{eqnarray}
\left| {\bf q};\hat{q}S\Lambda ;t\right\rangle ^{\pi a} & = &
\frac{1}{\sqrt{2}}(1-P_{12})\frac{1}{\sqrt{2}}(1+\eta _{\pi }P)\left|
{\bf q};\hat{q}S\Lambda \right\rangle \left| t\right\rangle \nonumber \\
 & = & \frac{1}{\sqrt{2}}(1-\eta _{\pi }(-)^{S+t})\left| t\right\rangle
\left| {\bf q};\hat{q}S\Lambda \right\rangle _{\pi }\label{8} 
\end{eqnarray}
To arrive at Eq.~(\ref{8}) we use well known properties of 
two-nucleon spin- and isospin states.

\noindent
In order to evaluate the normalization of the states in Eq.~(\ref{8}) 
 we need the
relation between \( \left| \hat{q}S\Lambda \right\rangle  \) 
and \( \left| -\hat{q}S\Lambda \right\rangle  \).
We use the definition of Eq.~(\ref{3}) for 
\( \left| -\hat{q}S\Lambda \right\rangle  \)
and the Wigner D-function, namely
\begin{eqnarray}
D_{\Lambda '\Lambda }^{S}(\hat{q}) & = & \left\langle \hat{z}S\Lambda '\right| R(\hat{q})\left| \hat{z}S\Lambda \right\rangle \nonumber \\
 & = & e^{-i\Lambda '\phi }d^{S}_{\Lambda '\Lambda }(\theta )\label{9}, 
\end{eqnarray}
to obtain
\begin{eqnarray}
\left| -\hat{q}S\Lambda \right\rangle  & = & \sum _{\Lambda '}D_{\Lambda '\Lambda }^{S}(-\hat{q})\left| \hat{z}S\Lambda '\right\rangle \nonumber \\
 & = & \sum _{\Lambda '}e^{-i(\phi +\pi )\Lambda '}d_{\Lambda '\Lambda }^{S}(\pi -\theta )\left| \hat{z}S\Lambda '\right\rangle \nonumber \\
 & = & \sum _{\Lambda '}e^{-i(\phi +\pi )\Lambda '}(-)^{S+\Lambda '}d_{\Lambda ',-\Lambda }^{S}(\theta )\left| \hat{z}S\Lambda '\right\rangle \nonumber \\
 & = & (-)^{S}\sum _{\Lambda '}D_{\Lambda ',-\Lambda }^{S}(\phi \theta 0)\left| \hat{z}S\Lambda '\right\rangle \nonumber \\
 & = & (-)^{S}\left| \hat{q}S-\Lambda \right\rangle \label{10} 
\end{eqnarray}
Now the normalization of the states given in Eq.~(\ref{8}) can be
worked out as
\begin{eqnarray}
 ^{\pi'a}\langle {\bf q}';\hat{q}'S'\Lambda ';t'| 
{\bf q};\hat{q}S\Lambda ;t \rangle ^{\pi a} 
 & =& \frac{1}{2}(1-\eta _{\pi '}(-)^{S'+t'})(1-\eta _{\pi}(-)^{S+t})
\delta _{t't}  \: _{\pi '}\langle {\bf q}';
\hat{q}'S'\Lambda ' |  {\bf q};\hat{q}S\Lambda \rangle _{\pi }
\nonumber \\
 & =  & (1-\eta _{\pi }(-)^{S+t})\delta _{t't}\delta _{\eta _{\pi '}\eta
_{\pi }}\delta _{S'S} \nonumber \\
 & & \left( \delta ({\bf q}'-{\bf q})\delta _{\Lambda
'\Lambda }+\eta _{\pi }(-)^{S}\delta ({\bf q}'+{\bf q})
 \delta _{\Lambda ',-\Lambda }\right) \label{11}
\end{eqnarray}

\noindent
Using this result it can also be verified that the completeness relation
of the states defined in Eq.~(\ref{8}) takes the form 
\begin{equation}
\label{12}
\sum _{S\Lambda \pi t}\int d^3 q  | {\bf q};\hat{q}S\Lambda; t\rangle ^{\pi a}
 \:\: \frac{1}{4} \:\: ^{\pi a}\langle {\bf q};\hat{q}S\Lambda ;t | = 1 .
\end{equation}

Equipped with the above given basis states one can formulate 
the Lippmann Schwinger integral equation. We
define as matrix elements
\begin{eqnarray}
T_{\Lambda '\Lambda }^{\pi St}({\bf q}',{\bf q}) & \equiv  & \left.
^{^{^{^{}}}}\right. ^{\pi a}\left\langle {\bf q}';\hat{q}'S\Lambda
';t\right| T\left| {\bf q};\hat{q}S\Lambda ;t\right\rangle ^{\pi a}\label{13} \\
V_{\Lambda '\Lambda }^{\pi St}({\bf q}',{\bf q}) & \equiv  & \left.
^{^{^{^{}}}}\right. ^{\pi a}\left\langle {\bf q}';\hat{q}'S\Lambda
';t\right| V\left| {\bf q};\hat{q}S\Lambda ;t\right\rangle ^{\pi a}\label{14} 
\end{eqnarray}
 Then using Eqs.~(\ref{12}-\ref{14}) the operator equation 
\( T=V+VG_{0}T \), which has a driving term the 
nucleon-nucleon (NN) potential V takes the form 
\begin{equation}
\label{15}
T_{\Lambda '\Lambda }^{\pi St}({\bf q}',{\bf q})=V_{\Lambda '\Lambda
}^{\pi St}({\bf q}',{\bf q})+\frac{1}{4}\sum _{\Lambda ''}\int d^3 q''
\, V_{\Lambda '\Lambda ''}^{\pi St}({\bf q}',{\bf
q}'')G_{0}(q'')T_{\Lambda ''\Lambda }^{\pi St}({\bf q}'',{\bf q})
\end{equation}

\noindent
We now distinguish between the two cases for total spin $S=0$ and $S=1$.
For  \( S=0 \) Eq.~(\ref{15}) is one equation, similar to the one
discussed  in Ref.\cite{9} for the bosonic case. 
For \( S=1 \) there are 3 coupled
equations to each \( \Lambda =-1,0,1 \). We use the property that $V$ conserves
spin and isospin, which is valid to a high degree of accuracy. The coupled
sets of equations in Eq.~(\ref{15}) can be further reduced as shown
below. Using Eq.~(\ref{8}) and the parity invariance of
$V$ one obtains 
\begin{eqnarray}
V_{\Lambda '\Lambda }^{\pi St}({\bf q}',{\bf q}) & = &
\frac{1}{2}(1-\eta _{\pi }(-)^{S+t})^{2}\left\langle t\right| \left.
_{_{_{_{_{}}}}}\right. _{\pi }\left\langle {\bf q}';\hat{q}'S\Lambda
'\right| V\left| {\bf q};\hat{q}S\Lambda \right\rangle _{\pi }\left| t\right\rangle \nonumber \\
 & = & \sqrt{2}(1-\eta _{\pi }(-)^{S+t})\left\langle t\right|
\left\langle \hat{q}'S\Lambda '\right| \left\langle {\bf q}'\right|
V\left| {\bf q}\right\rangle _{\pi }\left| \hat{q}S\Lambda \right\rangle
\left| t\right\rangle, \label{16} 
\end{eqnarray}
 where 
\begin{equation}
\label{17}
\left| {\bf q}\right\rangle _{\pi }\equiv \frac{1}{\sqrt{2}}\left(
\left| {\bf q}\right\rangle +\eta _{\pi }\left| -{\bf q}\right\rangle
\right). 
\end{equation}
 This expression can be connected to 
\( V_{-\Lambda '\Lambda }^{\pi St}({\bf q}',{\bf q}) \)
using Eqs.~(\ref{10}) and (\ref{17}) with the result 
\begin{equation}
\label{18}
V_{-\Lambda '\Lambda }^{\pi St}({\bf q}',{\bf q})=\eta _{\pi
}(-)^{S}V_{\Lambda '\Lambda }^{\pi St}(-{\bf q}',{\bf q}).
\end{equation}
 A corresponding relation is also valid for the matrix element 
\( T_{\Lambda '\Lambda }^{\pi St}({\bf q}',{\bf q}) \).
Similarly one finds 
\begin{equation}
\label{19}
V_{\Lambda ',-\Lambda }^{\pi St}({\bf q}',{\bf q})=\eta _{\pi
}(-)^{S}V_{\Lambda '\Lambda }^{\pi St}({\bf q}',-{\bf q}).
\end{equation}

\noindent
We can now simplify the set of coupled equations  (\ref{15}) for \( S=1 \) 
in the following way:
\begin{eqnarray}
\lefteqn {\int d^3 q''\, V_{\Lambda ',-1}^{\pi St}({\bf q}',{\bf
q}'')G_{0}(q'')T_{-1,\Lambda }^{\pi St}({\bf q}'',{\bf q})} &  & \nonumber \\
 &  & =\int d^3 q''\, \eta _{\pi }(-)^{S}V_{\Lambda ',1}^{\pi
St}({\bf q}',-{\bf q}'')G_{0}(q'')\eta _{\pi }(-)^{S}T_{1,\Lambda }^{\pi
St}(-{\bf q}'',{\bf q})\nonumber \\
 &  & =\int d^3 q''\, V_{\Lambda ',1}^{\pi St}({\bf q}',-{\bf
q}'')G_{0}(q'')T_{1,\Lambda }^{\pi St}(-{\bf q}'',{\bf q})\nonumber \\
 &  & =\int d^3 q''\, V_{\Lambda ',1}^{\pi
St}({\bf q}',{\bf q}'')G_{0}(q'')T_{1,\Lambda }^{\pi
St}({\bf q}'',{\bf q}).\label{20} 
\end{eqnarray}
This leads to 
\begin{eqnarray}
T_{\Lambda '\Lambda }^{\pi St}({\bf q}',{\bf q}) & = & V_{\Lambda
'\Lambda }^{\pi St}({\bf q}',{\bf q})+\frac{1}{2}\int d^3 q''\,
V_{\Lambda '1}^{\pi St}({\bf q}',{\bf q}'')G_{0}(q'')T_{1\Lambda }^{\pi
St}({\bf q}'',{\bf q})\nonumber \\
 &  & +\frac{1}{4}\int d^3 q''\, V_{\Lambda '0}^{\pi
St}({\bf q}',{\bf q}'')G_{0}(q'')T_{0\Lambda }^{\pi
St}({\bf q}'',{\bf q}).\label{21} 
\end{eqnarray}
We now have for the case $S=1$  two coupled equations, namely 
for \( \Lambda '=1,0 \) for each \( \Lambda  \).
Moreover due to Eq~(\ref{19}), which is also valid for the t-matrix $T$,
it is sufficient to consider \( \Lambda =0 \) and
$1$. As an aside it should be mentioned that because of the relation
$\pi (-)^{S+t}=-1$ the isospin quantum number $t$ is fixed, once $\pi$
and $S$ are chosen.

Finally we need to calculate  the physical T-matrix element expressed in
terms of the states 
\begin{equation}
\label{22}
\left| \nu _{1}\nu _{2}m_{1}m_{2}{\bf q}\right\rangle _{a}\equiv
\frac{1}{\sqrt{2}}(1-P_{12})\left| \nu _{1}\nu
_{2}m_{1}m_{2}{\bf q}\right\rangle ,
\end{equation}
 where \( \nu _{1}, \nu _{2} \) and \( m_{1},m_{2} \) are the magnetic isospin
and spin quantum numbers, \( {\bf q} \) describes the relative momentum of the
two nucleons, and $P_{12}$ is the permutation operator between the
particles $1$ and $2$. The physical T-matrix element is then given as 
\begin{equation}
 _a \langle \nu_1 \nu_2 m_1' m_2' {\bf q'}| T | 
\nu_1 \nu_2 m_1 m_2 {\bf q} \rangle _a 
 = \langle \nu_1\nu_2 m_1' m_2' {\bf q'} |
T(1-P_{12})| \nu_1 \nu_2 m_1 m_2 {\bf q}\rangle. \label{23} 
\end{equation}
The above equation now  has to be expressed in terms of 
\( T_{\Lambda '\Lambda }^{\pi St}({\bf q}',{\bf q}) \).
In doing so we  encounter the matrix element 
\begin{eqnarray}
\left\langle \hat{q}'S\Lambda '\right. \left| m_{1}m_{2}\right\rangle  & = & \sum _{S'}C(\frac{1}{2}\frac{1}{2}S';m_{1}m_{2}\Lambda _{0})\left\langle \hat{q}'S\Lambda '\right. \left| \hat{z}S'\Lambda _{0}\right\rangle \nonumber \\
 & = & C(\frac{1}{2}\frac{1}{2}S;m_{1}m_{2}\Lambda _{0})e^{i\Lambda _{0}\phi '}d^{S}_{\Lambda '\Lambda _{0}}(-\theta ')\nonumber \\
 & = & C(\frac{1}{2}\frac{1}{2}S;m_{1}m_{2}\Lambda _{0})e^{i\Lambda
_{0}\phi '}d^{S}_{\Lambda _{0}\Lambda '}(\theta '). \label{24} 
\end{eqnarray}
After some straightforward algebra one finds 
\begin{eqnarray}
\lefteqn { ^{\pi a}\langle {\bf q}';\hat{q}'S\Lambda ';t |
  \nu_1 \nu_2 m_1 m_2{\bf q}\rangle _a } 
&  & \nonumber \\
 & = & \frac{1}{\sqrt{2}}(1-\eta _{\pi}
(-)^{S+t})\frac{1}{\sqrt{2}} \langle t | \nu_1 \nu_2 \rangle 
 _{\pi }\langle {\bf q}';\hat{q}'S\Lambda'
 | (1-P_{12}) |  m_1 m_2 {\bf q} \rangle \nonumber \\
 & = & \frac{1}{\sqrt{2}}C(\frac{1}{2}\frac{1}{2}t;\nu_1\nu_2)
C(\frac{1}{2}\frac{1}{2}S;m_1 m_2 \Lambda_0)e^{i\Lambda_0\phi'}
d^S_{\Lambda_0\Lambda '}(\theta ') \nonumber \\
 &  & \quad (1-\eta _{\pi }(-)^{S+t})\left( \delta
({\bf q}'-{\bf q})+\eta _{\pi }\delta ({\bf q}'+{\bf q})\right). \label{25} 
\end{eqnarray}
We  then insert the completeness relation, Eq.~(\ref{12}), twice into
Eq.~(\ref{23}),
and using Eq.~(\ref{25}) obtain after some algebra 
\begin{eqnarray}
\lefteqn { _{a}\langle \nu _{1}\nu
_{2}m_{1}'m_{2}'{\bf q}'| T| \nu _{1}\nu _{2}m_{1}m_{2}{\bf q} \rangle _{a}}
 &  & \nonumber \\
 &  & =\frac{1}{4}e^{-i(\Lambda _{0}'\phi '-\Lambda _{0}\phi )}\sum _{S\pi t}(1-\eta _{\pi }(-)^{S+t})C(\frac{1}{2}\frac{1}{2}t;\nu _{1}\nu _{2})^{2}\nonumber \\
 &  & \quad C(\frac{1}{2}\frac{1}{2}S;m_{1}'m_{2}'\Lambda _{0}')C(\frac{1}{2}\frac{1}{2}S;m_{1}m_{2}\Lambda _{0})\nonumber \\
 &  & \quad \sum _{\Lambda '\Lambda }d^{S}_{\Lambda _{0}'\Lambda
'}(\theta ')d^{S}_{\Lambda _{0}\Lambda }(\theta )T_{\Lambda '\Lambda
}^{\pi St}({\bf q}',{\bf q}). \label{26} 
\end{eqnarray}
For our calculation we  choose \( \hat{q}=\hat{z} \) and use 
$d_{\Lambda _{0}\Lambda }^{S}(0)=\delta _{\Lambda _{0}\Lambda }$.
Furthermore,  as will be shown in Section III, the azimuthal dependence
of ${\bf q}'$ and ${\bf q}$ can be factored out and one finds 
\begin{equation}
\label{27}
T_{\Lambda '\Lambda }^{\pi St}({\bf q}',{\bf q})=e^{i\Lambda (\phi '-\phi )}T_{\Lambda '\Lambda }^{\pi St}(q',q,\theta ')
\end{equation}
With this Eq.~(\ref{26}) can be written as (note $ q'=q $) 
\begin{eqnarray}
\lefteqn {\left. _{_{_{_{_{}}}}}\right. _{a}\left\langle \nu _{1}\nu
_{2}m_{1}'m_{2}'q\hat{q}'\right| T\left| \nu _{1}\nu _{2}m_{1}m_{2}{\bf q}\right\rangle _{a}} &  & \nonumber \\
 &  & =\frac{1}{4}e^{-i(\Lambda _{0}'-\Lambda _{0})\phi '}\sum _{S\pi t}C(\frac{1}{2}\frac{1}{2}t;\nu _{1}\nu _{2})^{2}(1-\eta _{\pi }(-)^{S+t})\nonumber \\
 &  & \quad C(\frac{1}{2}\frac{1}{2}S;m_{1}'m_{2}'\Lambda _{0}')C(\frac{1}{2}\frac{1}{2}S;m_{1}m_{2}\Lambda _{0})\sum _{\Lambda '}d^{S}_{\Lambda _{0}'\Lambda '}(\theta ')T_{\Lambda '\Lambda _{0}}^{\pi St}(q,q,\theta ')\label{28} 
\end{eqnarray}

\section{General structure of the potential operator and final form of the
scattering  equation}

As it is well known, rotational-, parity-, and time reversal invariance 
restricts any NN potential V to be formed out of six terms
\cite{wolfenstein}, which are given by 
\begin{eqnarray}
W_{1} & = & 1\label{29} \nonumber \\
W_{2} & = & ({\ffat \sigma }_{1}+{\ffat \sigma }_{2})\cdot \hat{N}\label{30}
\nonumber \\
W_{3} & = & {\ffat \sigma }_{1}\cdot \hat{N}\, {\ffat \sigma }_{2}\cdot \hat{N}\label{31} 
\nonumber \\
W_{4} & = & {\ffat \sigma }_{1}\cdot \hat{P}\, {\ffat \sigma }_{2}\cdot \hat{P}\label{32} 
\nonumber \\
W_{5} & = & {\ffat \sigma }_{1}\cdot \hat{K}\, {\ffat \sigma }_{2}\cdot \hat{K}\label{33} 
\nonumber \\
W_{6} & = & {\ffat \sigma }_{1}\cdot \hat{P}\, {\ffat \sigma }_{2}\cdot
\hat{K}+{\ffat \sigma }_{1}\cdot \hat{K}\, {\ffat \sigma }_{2}\cdot \hat{P},
 \label{34} 
\end{eqnarray}
 where the unit vectors \( \hat{N},\, \hat{P} \) and \( \hat{K} \) are defined
in terms of the relative momenta \( {\bf q} \) and \( {\bf q}' \) as 
\begin{eqnarray}
\hat{N} & = & \frac{({\bf q}\times {\bf q}')}{|{\bf q}\times {\bf q}'|}\label{35} \\
\hat{P} & = & \frac{({\bf q}'+{\bf q})}{|{\bf q}'+{\bf q}|}\label{36} \\
\hat{K} & = & \frac{({\bf q}'-{\bf q})}{|{\bf q}'-{\bf q}|}.\label{37} 
\end{eqnarray}
 The most general potential is then given as linear combination of
those operators, namely 
\begin{equation}
\label{38}
 \langle {\bf q'}|V| {\bf q} \rangle \equiv 
V({\bf q'},{\bf q})=\sum _{i=1}^{6}v_{i}(q',q,\gamma) W_{i},
\end{equation}
 where $ v_{i}(q',q,\gamma)$  are scalar functions depending on 
the magnitudes of the vectors ${\bf q'}$, ${\bf q}$, and 
the angle between the two, $\gamma \equiv \hat{q}'\cdot \hat{q} $.
In Section IV we will provide specific realizations of the potential 
$V({\bf q'},{\bf q})$.

The evaluation of the matrix elements of Eq.~(\ref{14}) for the set of operators
\( W_{i} \) is in principle possible. However,  it is simpler  to define a different
set of six independent operators, $\Omega _{i}$, which is more adapt to
our choice of basis states. We introduce the following set of operators
\begin{eqnarray}
\Omega _{1} & = & 1\label{39} \nonumber  \\
\Omega _{2} & = & {\bf S}^{2}\label{40} \nonumber  \\
\Omega _{3} & = & {\bf S}\cdot \hat{q}'\, {\bf S}\cdot
\hat{q}'\label{41} \nonumber  \\
\Omega _{4} & = & {\bf S}\cdot \hat{q}'\, {\bf S}\cdot \hat{q}\label{42}
\nonumber  \\
\Omega _{5} & = & ({\bf S}\cdot \hat{q}')^{2}\, ({\bf S}\cdot
\hat{q})^{2}\label{43}  \nonumber \\
\Omega _{6} & = & {\bf S}\cdot \hat{q}\, {\bf S}\cdot \hat{q}\label{44}. 
\end{eqnarray}
The two sets of operators given in  Eqs.~(\ref{34}) and (\ref{44}) can be easily 
related via
\begin{equation}
\label{45}
W_{i}=\sum _{j}A_{ij}\Omega _{j}.
\end{equation}
 The transformation matrix $A_{ij}$ and its inverse are explicitly 
given in Appendix A. 

Using the operators $\Omega_i$, 
the spin-dependent part of the matrix elements of Eq.~(\ref{14}) 
can then be reduced
to the evaluation of  matrix elements of the type 
$ \left\langle \hat{q}'S\Lambda '\right| \Omega _{i}\left| 
\hat{q}S\Lambda \right\rangle $,
namely
\begin{eqnarray}
\left\langle \hat{q}'S\Lambda '\right| \Omega _{1}\left| \hat{q}S\Lambda
\right\rangle  & = & \left\langle \hat{q}'S\Lambda '\right. \left|
\hat{q}S\Lambda \right\rangle \label{48} \nonumber \\
\left\langle \hat{q}'S\Lambda '\right| \Omega _{2}\left| \hat{q}S\Lambda
\right\rangle  & = & S(S+1)\left\langle \hat{q}'S\Lambda '\right. \left|
\hat{q}S\Lambda \right\rangle \label{49} \nonumber \\
\left\langle \hat{q}'S\Lambda '\right| \Omega _{3}\left| \hat{q}S\Lambda
\right\rangle  & = & \Lambda '^{2}\left\langle \hat{q}'S\Lambda '\right.
\left| \hat{q}S\Lambda \right\rangle \label{50} \nonumber \\
\left\langle \hat{q}'S\Lambda '\right| \Omega _{4}\left| \hat{q}S\Lambda
\right\rangle  & = & \Lambda '\Lambda \left\langle \hat{q}'S\Lambda
'\right. \left| \hat{q}S\Lambda \right\rangle \label{51} \nonumber \\
\left\langle \hat{q}'S\Lambda '\right| \Omega _{5}\left| \hat{q}S\Lambda
\right\rangle  & = & \Lambda '^{2}\Lambda ^{2}\left\langle
\hat{q}'S\Lambda '\right. \left| \hat{q}S\Lambda \right\rangle
\label{52} \nonumber \\
\left\langle \hat{q}'S\Lambda '\right| \Omega _{6}\left| \hat{q}S\Lambda
\right\rangle  & = & \Lambda ^{2}\left\langle \hat{q}'S\Lambda '\right.
\left| \hat{q}S\Lambda \right\rangle . 
\label{53} 
\end{eqnarray}
Now we are left with determining the overlap of the states defined in Eq.~(\ref{3}). 
Using Eq.~(\ref{9}) this gives
\begin{eqnarray}
\left\langle \hat{q}'S\Lambda '\right. \left| \hat{q}S\Lambda \right\rangle  & = & \sum _{M}\left\langle \hat{q}'S\Lambda '\right. \left| \hat{z}SM\right\rangle \left\langle \hat{z}SM\right. \left| \hat{q}S\Lambda \right\rangle \nonumber \\
 & = & \sum _{M}D^{S}_{M\Lambda '}(\phi '\theta '0)^{\ast }D^{S}_{M\Lambda }(\phi \theta 0)\nonumber \\
 & = & \sum ^{S}_{M=-S}e^{iM(\phi '-\phi )}d^{S}_{M\Lambda '}(\theta ')d^{S}_{M\Lambda }(\theta )\label{54} 
\end{eqnarray}
A special case is the situation where $\bf q$ 
points in $z$-direction, where one
obtains the simple form 
\begin{equation}
\label{55}
\left\langle \hat{q}'S\Lambda '\right. \left| \hat{z}S\Lambda \right\rangle =e^{i\Lambda (\phi '-\phi )}d^{S}_{\Lambda '\Lambda }(\theta ')
\end{equation}

With these preliminaries
the potential matrix element of Eq.~(\ref{14}) can be written as
\begin{eqnarray}
V_{\Lambda '\Lambda }^{\pi St}({\bf q}',{\bf q}) & = & \sqrt{2}(1-\eta _{\pi }(-)^{S+t})\nonumber \\
 &  & \sum _{ij}A_{ij}\left\langle \hat{q}'S\Lambda '\right| \Omega
_{j}\left| \hat{q}S\Lambda \right\rangle \left\langle t\right|
\left\langle {\bf q}'\right| v_{i}\left| {\bf q}\right\rangle _{\pi }\left| t\right\rangle \label{56} 
\end{eqnarray}
The coefficients $ A_{ij} $ and the matrix elements 
$\left\langle {\bf q}'\right| v_{i}\left| {\bf q}\right\rangle _{\pi }$ 
depend on the angle between ${\bf q}$ and ${\bf q'}$ and thus on $\gamma$. 
Therefore their azimuthal dependence is determined by
$\cos (\phi '-\phi )$. Furthermore, the matrix elements 
$\Omega _{j}$ depend
on the angles \( \phi ' \) and \( \phi  \) as shown in the Eq.~(\ref{54}).
Thus we can schematically indicate 
the azimuthal dependence  of the potential matrix elements as 
\begin{equation}
\label{57}
V_{\Lambda '\Lambda }^{\pi St}({\bf q}',{\bf q}) \equiv 
V_{\Lambda '\Lambda }^{\pi St}
\left\{ e^{iM(\phi '-\phi )},\cos(\phi '-\phi) \right\} 
\end{equation}
 For the special case $\hat{q}=\hat{z}$ this reduces to the simpler form
\begin{equation}
\label{58}
V_{\Lambda '\Lambda }^{\pi St}({\bf q}',{\bf q})=
e^{i\Lambda (\phi '-\phi )}V_{\Lambda '\Lambda }^{\pi St}(q',q,\theta '),
\end{equation}
which  is the driving term in the coupled set of Eqs.~(\ref{21}).

We assume that this simple dependence on \( \phi ' \) and \( \phi  \) given
in (\ref{58}) carries over to the solution of the integral equation
 and choose as ansatz 
\begin{equation}
\label{59}
T_{\Lambda '\Lambda }^{\pi St}({\bf q}',{\bf q})=e^{i\Lambda (\phi '-\phi )}T_{\Lambda '\Lambda }^{\pi St}(q',q,\theta ')
\end{equation}
 Inserting this into Eq.~(\ref{15}) one obtains 
\begin{eqnarray}
T_{\Lambda '\Lambda }^{\pi St}({\bf q}',q\hat{z}) & = & e^{i\Lambda (\phi '-\phi )}V_{\Lambda '\Lambda }^{\pi St}(q',q,\theta ')\nonumber \\
 &  & +\frac{1}{4}\sum _{\Lambda ''}\int d^3 q''\, 
V_{\Lambda '\Lambda }^{\pi St} \left\{ e^{iM(\phi '-\phi '')},\cos (\phi '-\phi
''),q',q'' \right\} \nonumber \\
 &  & \qquad \qquad G_{0}(q'')e^{i\Lambda (\phi ''-\phi )}
T_{\Lambda ''\Lambda }^{\pi St}(q'',q,\theta '')\nonumber \\
 & = & e^{i\Lambda (\phi '-\phi )}V_{\Lambda '\Lambda }^{\pi St}(q',q,\theta ')
\nonumber \\
 &  & +\frac{1}{4}e^{i\Lambda (\phi '-\phi )}\sum _{\Lambda ''}
\int ^{\infty }_{0}dq''\, q''^{2}\int ^{\pi }_{0}d\theta ''\, \sin \theta ''
\int ^{2\pi }_{0}d\phi ''\nonumber \\
 &  & \qquad V_{\Lambda '\Lambda }^{\pi St}\left\{ e^{iM(\phi '-\phi '')},
\cos (\phi '-\phi ''),q',q''\right\} G_{0}(q'')\nonumber \\
 &  & \qquad e^{i\Lambda (\phi ''-\phi ')}T_{\Lambda ''\Lambda }^{\pi St}(q'',q,\theta '')\label{60} 
\end{eqnarray}
 The integrand is periodical with respect to  $\phi''$, 
with the period being \( 2\pi  \). Consequently one can set \( \phi '=0 \) for the
$\phi''$-integration. This leads to
\begin{eqnarray}
T_{\Lambda '\Lambda }^{\pi St}({\bf q}',q\hat{z}) & = & e^{i\Lambda
(\phi '-\phi )}\Biggl[ V_{\Lambda '\Lambda }^{\pi St}(q',q,\theta ') \nonumber \\
 &  & +\frac{1}{4}\sum _{\Lambda ''}\int ^{\infty }_{0}dq''\, q''^{2}
\int ^{\pi }_{0}d\theta ''\, \sin \theta ''\int ^{2\pi }_{0}d\phi ''\nonumber \\
 &  & \qquad V_{\Lambda '\Lambda }^{\pi St}\left\{ e^{-iM\phi ''},\cos \phi
'',q',q'' \right\} G_{0}(q'')e^{i\Lambda \phi ''}
T_{\Lambda ''\Lambda }^{\pi St}(q'',q,\theta '') \Biggr] \nonumber \\
 & \equiv  & e^{i\Lambda (\phi '-\phi )}T_{\Lambda '\Lambda }^{\pi St}(q',q,\theta'),
\label{61} 
\end{eqnarray}
verifying the correctness of the ansatz of Eq.~(\ref{59}).

Inserting this result into Eq.~(\ref{21}) gives
\begin{eqnarray}
T_{\Lambda '\Lambda }^{\pi St}(q',q,\theta ') & = & 
V_{\Lambda '\Lambda }^{\pi St}(q',q,\theta ')\nonumber \\
 &  & +\frac{1}{2}\int ^{\infty }_{0}dq''\, q''^{2}
\int ^{1}_{-1}d(\cos \theta '')\nonumber \\
 &  & \qquad v_{\Lambda '1}^{\pi St,\Lambda }(q',q'',\theta ',\theta '')
G_{0}(q'')T_{1\Lambda }^{\pi St}(q'',q,\theta '')\nonumber \\
 &  & +\frac{1}{4}\int ^{\infty }_{0}dq''\, q''^{2}\int ^{1}_{-1}d(\cos \theta '')
\nonumber \\
 &  & \qquad v_{\Lambda '0}^{\pi St,\Lambda }(q',q'',\theta ',\theta '')G_{0}(q'')
T_{0\Lambda }^{\pi St}(q'',q,\theta ''), \label{62} 
\end{eqnarray}
 with 
\begin{equation}
\label{63}
v_{\Lambda '\Lambda ''}^{\pi St,\Lambda }(q',q'',\theta ',\theta
'')\equiv \int ^{2\pi }_{0}d\phi ''e^{-i\Lambda (\phi '-\phi
'')}V_{\Lambda '\Lambda ''}^{\pi St}({\bf q}',{\bf q}'')
\end{equation}
 The driving term of Eq.~(\ref{62}) is a special case of Eq.~(\ref{63}) for 
$\theta ''=0$ and $\Lambda ''=\Lambda$
(up to a factor \( (2\pi )^{-1} \)). 
In summary,  for the case \( S=1 \) we end up with two
coupled integral equations in two variables for given \( \Lambda  \)-values
(only \( \Lambda =1,0 \) are necessary). The azimuthal integration over
\( \phi '' \) can be performed independently, and does not enter 
the integral kernel.

The case \( S=0 \) is much simpler, one has only one integral equation in two
variables, similar in its structure to
the two-boson case discussed in Ref.~\cite{9}.

\section{Representation of NN potentials}

In this Section we want to demonstrate the relative ease with which modern 
NN potentials that are given in operator form
can be incorporated in a three-dimensional formalism. Our choices of NN potentials
are a Bonn one-boson-exchange (OBE) potential \cite{physrep} in the parameterization
of Bonn-B \cite{16} and the Argonne coordinate
space potential AV18 \cite{av18}.

The OBE potential consists of  pseudoscalar-, scalar-, and vector meson
exchanges,  derived from the corresponding Feynman diagrams. A three-dimensional
reduction of the Bethe-Salpeter equation is as achieved via the 
Blankenbecler-Sugar reduction. Details can be found in Refs. \cite{16,physrep}.
The resulting potential operators are given as
\begin{eqnarray}
V^{s}({\bf q'},{\bf q}) & = & -\frac{g_s^2}{(2\pi)^3} \sqrt{\frac{m}{E'}}
  \sqrt{\frac{m}{E}} \:
  \bar{u}({\bf q}')u({\bf q})\bar{u}(-{\bf q}')u(-{\bf q}) \:
 \frac{F_s[({\bf q'}-{\bf q})^2]} {({\bf q'}-{\bf q})^{2}+\mu _{s}^{2}}
   \label{64} \\
 &  & \nonumber \\
V^{ps}({\bf q'},{\bf q}) & = & \frac{g_{ps}^2}{(2\pi)^3} \sqrt{\frac{m}{E'}}
\sqrt{\frac{m}{E}} \:
\bar{u}({\bf q}')\gamma^5 u({\bf q})\bar{u}(-{\bf q}')\gamma^5 u(-{\bf q}) \:
\frac{F_{ps}[({\bf q'}-{\bf q})^2]}
{({\bf q}'-{\bf q})^2+\mu _{ps}^2}  \label{65} \\
 &  & \nonumber \\
V^{v}({\bf q'},{\bf q}) & = & \frac{F_v[({\bf q'}-{\bf q})^2]}
{({\bf q}'-{\bf q})^{2}+\mu _{v}^{2}} \; \sqrt{\frac{m}{E'}} \sqrt{\frac{m}{E}} \:
\frac{1}{(2\pi)^3}
\left[ g_{v}^{2}\bar{u}({\bf q}')\gamma ^{\mu }u({\bf q})\bar{u}(-{\bf
q}')\gamma _{\mu }u(-{\bf q})\right. \label{66} \\
 &  & +\frac{f_{v}^{2}}{4m^{2}}\left\{ 4m^{2}\, \bar{u}({\bf q}')\gamma
^{\mu }u({\bf q})\bar{u}(-{\bf q}')\gamma _{\mu }u(-{\bf q})\right. \nonumber \\
 &  & \qquad -2m\, \bar{u}({\bf q}')\gamma ^{\mu }u({\bf
q})\bar{u}(-{\bf q}')[(E'-E)(g_{\mu }^{0}-\gamma _{\mu }\gamma
^{0})+(p_{2}+p_{2}')_{\mu }]u(-{\bf q})\nonumber \\
 &  & \qquad -2m\, \bar{u}({\bf q}')[(E'-E)(g^{0\mu }-\gamma ^{\mu
}\gamma ^{0})+(p_{1}+p_{1}')^{\mu }]u({\bf q})\bar{u}(-{\bf q}')\gamma
_{\mu }u(-{\bf q})\nonumber \\
 &  & \qquad +\bar{u}({\bf q}')[(E'-E)(g^{0\mu }-\gamma ^{\mu }\gamma
^{0})+(p_{1}+p_{1}')^{\mu }]u({\bf q})\nonumber \\
 &  & \qquad \left. \bar{u}(-{\bf q}')[(E'-E)(g_{\mu }^{0}-\gamma _{\mu
}\gamma ^{0})+(p_{2}+p_{2}')_{\mu }]u(-{\bf q})\right\} \nonumber \\
 &  & +\frac{g_{v}f_{v}}{2m}\left\{ 4m\, \bar{u}({\bf q}')\gamma ^{\mu
}u({\bf q})\bar{u}(-{\bf q}')\gamma _{\mu }u(-{\bf q})\right. \nonumber \\
 &  & \qquad -\bar{u}({\bf q}')\gamma ^{\mu }u({\bf q})\bar{u}(-{\bf
q}')[(E'-E)(g_{\mu }^{0}-\gamma _{\mu }\gamma ^{0})+(p_{2}+p_{2}')_{\mu
}]u(-{\bf q})\nonumber \\
 &  & \qquad -\left. \left. \bar{u}({\bf q}')[(E'-E)(g^{0\mu }-\gamma
^{\mu }\gamma ^{0})+(p_{1}+p_{1}')^{\mu }]u({\bf q})\bar{u}(-{\bf
q}')\gamma _{\mu }u(-{\bf q})\right\} \right]. \nonumber 
\end{eqnarray}
Here $m$ stands for the nucleon mass.
 In the case of the vector potential one has
$(p_{1}+p_{1}')^{\mu }=(E+E',{\bf q}+{\bf q}')$
 and 
$(p_{2}+p_{2}')^{\mu }=(E+E',-{\bf q}-{\bf q}')$.
 The  coupling constants $g_{ps,s,v}$ and $f_v$, the cutoff functions $F$ and
the meson masses are  given in \cite{16}.

In order to bring this OBE potential in a form consistent with our 
three-dimensional (3D) equations, the bilinear Dirac forms have to be expressed
in terms of the operators $W_i$ of Eqs. (\ref{34}). 
The result in form of the Wolfenstein operators of Eq.~(\ref{38}) is given
 in Appendix B.

Another often used modern NN potential is the Argonne potential AV18
\cite{av18}. It is originally presented in configuration space and has the
general form 
\begin{equation}
\label{69}
V(NN) = v^{EM}(NN) + v^{\pi}(NN) + v^R(NN).
\end{equation}
Here $v^{EM}(NN)$ represents an electromagnetic part, which we omit in
this work. The one-pion-exchange (OPE) part $v^{\pi}(NN)$ is charge
dependent and has the standard form. The Yukawa and tensor functions
contain exponential cutoffs, thus do not have an analytical Fourier
transform to momentum space. The intermediate- and short-range
phenomenological part $v^R(NN)$ is expressed as a sum of central, tensor,
spin-orbit, $L^2$, and quadratic spin-orbit pieces (abbreviated as
$c, t, ls, l2, ls2$, respectively) in different spin ($S$) and iso-spin
($T$) states:
\begin{equation}
\label{70}
v^R_{ST}(NN)= v^c_{ST}(r) {\bf 1} + v^t_{ST}(r) S_{12} + v^{ls}_{ST}(r)
{\bf L}\cdot{\bf S} + v^{l2}_{ST}(r) L^2 + 
v^{ls2}_{ST}(r)({\bf L}\cdot{\bf S})^2,
\end{equation}
where $S_{12}$ denotes the standard tensor operator. The specific form of
the radial functions as well as the potential parameters are given in
Ref.~\cite{av18}. For applying this potential in our formulation, we need
to perform the transition to momentum space. For the terms contributing to 
$v^R(NN)$ we obtain explicitly
\begin{eqnarray}
v^c_{ST}({\bf q}',{\bf q}) & = & \frac{1}{2\pi ^{2}}\int _{0}^{\infty }dr
r^{2}j_{0}(\rho r)\, v^c_{ST}(r)\label{71} \\
 &  & \nonumber \\
v^t_{ST}({\bf q}',{\bf q}) & = & \frac{1}{2\pi ^{2}}
\left( \frac{-3 \: {\ffat \sigma_1}\cdot ({\bf q'}-{\bf q})  \: {\ffat \sigma_2}
\cdot ({\bf q'}-{\bf q})}
 {\rho^2}
+ {\ffat \sigma_1}\cdot {\ffat \sigma_2} \right)
\int _{0}^{\infty }dr r^2 j_2 (\rho r)v^t_{ST}(r)\label{72} \\
 &  & \nonumber \\
v^{ls}_{ST}({\bf q}',{\bf q}) & = & \frac{i}{2\pi^2}{\bf S}\cdot 
({\bf q}\times {\bf q}')\frac{1}{\rho }\int _{0}^{\infty }dr\, r^3 j_{1}(\rho r)
v^{ls}_{ST}(r)\label{73} \\
 &  & \nonumber \\
v^{l2}_{ST}({\bf q}',{\bf q}) & = & \frac{1}{2\pi ^{2}}{\bf q}\cdot {\bf q'}
\frac{2}{\rho }\int _{0}^{\infty }dr\, r^3 j_{1}(\rho r)v^{l2}_{ST}(r)
\nonumber \\
 &  & -\frac{1}{2\pi ^{2}}\left[ {\bf q}'^{2}{\bf q}^{2}(1-\gamma ^{2})\right] 
\frac{1}{\rho ^{2}}\int _{0}^{\infty }dr\, r^4 j_2(\rho r)v^{l2}_{ST}(r)
\label{74} \\
 &  & \nonumber \\
v^{ls2}_{ST}({\bf q}',{\bf q}) & = & \frac{1}{2\pi ^{2}}\left( {\bf S}^{2}{\bf
q}\cdot {\bf q}'-{\bf S}\cdot {\bf q} {\bf S}\cdot {\bf q'}\right) 
\frac{1}{\rho} \int _{0}^{\infty }dr\, r^3 j_1(\rho r)v^{ls2}_{ST}(r)
\nonumber \\
 &  & -\frac{1}{2\pi ^{2}}\left( \frac{1}{2}({\bf q}\times {\bf
q'})^2+\frac{1}{2}{\ffat \sigma }_{1}\cdot ({\bf q}\times {\bf q}')
{\ffat \sigma }_{2}\cdot ({\bf q}\times {\bf q'})\right) \nonumber \\
 &  & \frac{1}{\rho^2}\int _{0}^{\infty }dr\, r^4 j_{2}(\rho
r)v^{ls2}_{ST}(r), \label{75} 
\end{eqnarray}
 where  $\rho \equiv |{\bf q'}-{\bf q}|$. 
The resulting operators can be
easily represented as function of the Wolfenstein operators $W_i$. The
final expression are given in Appendix C.

\section{Connection to a partial wave representation}

In order to compare with standard partial representations, especially 
NN phase shifts, we need to make connection to the standard partial wave
representation.
The partial wave projected T-matrix element is defined as 
\begin{equation}
\label{79}
T_{l'l}^{Sjt}(q)\equiv \left\langle q(l'S)jmtm_{t}\right| T\left|
q(lS)jmtm_{t}\right\rangle ,
\end{equation}
where the states $ \left| q(lS)jmtm_{t}\right\rangle$ are given as 
\begin{equation}
\label{80}
\left| q(lS)jmtm_{t}\right\rangle =\sum _{\mu }C(lSj,\mu m-\mu )\left|
ql\mu \right\rangle \left| Sm-\mu \right\rangle \left|
tm_{t}\right\rangle. 
\end{equation}
We choose the standard normalization for those states, namely
\begin{eqnarray}
\left\langle q'(l'S')j'm't'm_{t}'\right. \left| q(lS)jmtm_{t}\right\rangle
& = & \frac{\delta (q-q)}{q'q}\delta _{l'l}\delta _{S'S}\delta
_{j'j}\delta _{m'm}\delta _{t't}\delta _{m_{t}'m_{t}}. \label{81} 
\end{eqnarray}

The connection to the states $\left| {\bf q};\hat{q}S\Lambda \right\rangle$
can be  found using Eq.~(\ref{9}), and one finds
\begin{eqnarray}
\left| {\bf q};\hat{q}S\Lambda \right\rangle  & = & 
\sum _{ljm}\left| q(lS)jm\right\rangle \nonumber \\
 &  & \sum _{\mu }C(lSj;\mu m-\mu )Y_{l\mu }^{\ast }(\hat{q})e^{-i(m-\mu
)\phi }d^{S}_{m-\mu ,\Lambda }(\theta ).\label{82} 
\end{eqnarray}
Consequently one obtains for the T-matrix
\begin{eqnarray}
T_{\Lambda '\Lambda }^{\pi St}({\bf q}',{\bf q}) & = & \sqrt{2}(1-\eta
_{\pi }(-)^{S+t})\left. _{_{_{_{_{}}}}}\right. _{\pi }\left\langle {\bf
q}';\hat{q}'S\Lambda '\right| \left\langle t\right| T\left|
t\right\rangle \left| {\bf q}\right\rangle \left| \hat{q}S\Lambda \right\rangle 
\nonumber \\
 &= & \frac{1}{2}(1-\eta _{\pi }(-)^{S+t})\sum _{l'ljm}T_{l'l}^{Sjt}(q)(1+\eta _{\pi }(-)^{l'})(1+\eta _{\pi }(-)^{l})\nonumber \\
 &  & \quad \sum _{\mu '}C(l'Sj;\mu ',m-\mu ')Y_{l'\mu '}(\hat{q}')e^{i(m-\mu ')\phi '}d^{S}_{m-\mu ',\Lambda '}(\theta ')\nonumber \\
 &  & \quad \sum _{\mu }C(lSj;\mu ,m-\mu )Y^{\ast }_{l\mu
}(\hat{q})e^{-i(m-\mu )\phi }d^{S}_{m-\mu ,\Lambda }(\theta ). \label{83} 
\end{eqnarray}
For ${\bf q}$ parallel to the z-axis, i.e.  $\hat{q}=\hat{z}$, one finds
after some straightforward
algebra the on-shell relation 
\begin{eqnarray}
T_{\Lambda '\Lambda }^{\pi St}(q,q,\theta') & = & 
\frac{1}{2}(1-\eta _{\pi }(-)^{S+t})\sum _{l'lj}T_{l'l}^{Sjt}(q)
(1+\eta _{\pi }(-)^{l'})(1+\eta _{\pi }(-)^{l})\nonumber \\
& & \sqrt{\frac{2l'+1}{4\pi }}C(l'Sj;0\Lambda ')d_{\Lambda \Lambda'}^j(\theta')
 \sqrt{\frac{2l+1}{4\pi }}C(lSj;0\Lambda ). \label{85} 
\end{eqnarray}
Here the relation 
\begin{equation}
\label{84}
Y_{l'\mu'}(\hat{q}')= \sqrt{\frac{2l'+1}{4\pi }}D^{l'\ast}_{\mu',0}(\phi'\theta'0)=
\sqrt{\frac{2l'+1}{4\pi }}(-)^{\mu '}D^{l'}_{-\mu ',0} (\phi '\theta '0)
\end{equation}
was used together with an addition theorem for D-functions.

Once the  partial wave projected T-matrices are calculated, we can use the
the well known connection to the partial wave S-matrices 
\begin{equation}
\label{86}
S^{Sjt}_{l'l}=\delta _{l'l}-\pi imqT_{l'l}^{Sjt},
\end{equation}
which are parameterized by the standard partial wave phase shifts.

The inversion of Eq.~(\ref{85}) can be accomplished by using the orthogonality
relation 
\begin{equation}
\label{87}
\int _{0}^{2\pi }d\phi \int _{0}^{\pi }d\theta \, \sin \theta D_{\mu
m}^{j_{1}\ast }(\phi \theta 0)D_{\mu m}^{j_{2}}(\phi \theta 0)=\frac{4\pi
}{2j_{1}+1}\delta _{j_{1}j_{2}}.
\end{equation}
In addition we take into account  Eq.~(\ref{59}) for the on-shell T-matrix and
find after a few steps 
\begin{eqnarray}
T_{l'l}^{Sjt}(q) & = & 4\pi (1-\eta _{\pi }(-)^{S+t})^{-1}(1+\eta _{\pi }(-)^{l'})^{-1}(1+\eta _{\pi }(-)^{l})^{-1}\nonumber \\
 &  & \frac{\sqrt{2l'+1}\sqrt{2l+1}}{2j+1}\sum _{\Lambda '\Lambda }C(l'Sj;0\Lambda ')C(lSj;0\Lambda )\nonumber \\
 &  & \int _{-1}^{1}d(\cos \theta ')d_{\Lambda \Lambda '}^{j}(\theta
')T^{\pi St}_{\Lambda '\Lambda }(q,q,\theta '). \label{88} 
\end{eqnarray}

Finally we connect the physical T-matrix element to the partial wave projected S-matrices.
We insert (\ref{85}) into (\ref{28}). In proceeding we use
\begin{eqnarray}
 D^{j*}_{\Lambda \Lambda'} (\phi' \theta'0) =
(-)^{\Lambda - \Lambda'} e^{i \Lambda \phi'}
 d^j_{- \Lambda -\Lambda'} (\theta')
\end{eqnarray}
and an addition theorem of $D$-functions together with (\ref{84}). Then we end up with the 
standard form
\begin{eqnarray}
\lefteqn{\: _{a} \langle \nu_{1} \nu_{2} m_{1}' m_{2}' q {\hat q}' | T
| \nu_{1} \nu_{2} m_{1} m_{2} {\vec q} \rangle _{a}}\nonumber\\
&& =  \frac{1}{2} {1 \over \sqrt{4 \pi}} {1 \over \pi im q} 
\sum_{St} C({1\over 2} {1\over 2} t, \nu_1 \nu_2)^{2}
C({1\over 2} {1\over 2}S, m_{1}'m_{2}' \Lambda_{0}')
C({1\over 2} {1\over 2}S, m_{1} m_{2} \Lambda_{0})\nonumber\\
&& \quad \sum_{jl'l} (1 - (-)^{l+S+t})
(1-(-)^{l'+S+t}) (\delta_{ll'} - S_{l'l}^{Sjt})\nonumber\\
&& \quad \sqrt{2l + 1} C(lSj, 0\Lambda_{0}) C(l'Sj, \Lambda_{0} - \Lambda_{0}', \Lambda_{0}')
Y_{l', \Lambda_{0} - \Lambda_{0}'}(\theta',\phi')  \label{eq:5.12}
\end{eqnarray}
One can choose $\phi' = 0$ by putting ${\bf q}'$ into the $x-z$ plane

In order to unambiguously define the normalization we also give the expression 
for the spin-averaged differential cross section for nucleon species 
$\nu_{1} \nu_{2}$ as
\begin{equation}
\frac{d\sigma}{d\Omega} = (2\pi)^4 \left(\frac{m}{2}\right)^2 \frac{1}{4}
\sum_{m'_1 m'_2 m_1 m_2}  |_a\langle \nu_{1} \nu_{2} m_{1}' m_{2}' q {\hat q}'
 | T | \nu_{1} \nu_{2} m_{1} m_{2} {\bf q} \rangle_a |^2
\label{eq:5.13}
\end{equation}

\section{Results and Discussion}

In order to demonstrate the feasibility of our formulation when 
applied to NN scattering, we present numerical results for two NN potentials
of quite different character, the Bonn-B \cite{16} and the AV18 \cite{av18} potential models. 
For calculating the NN t-matrix we solve the integral equation, Eq.~(\ref{62}).
For S=0 this is a single, two-dimensional integral equation in two
variables, for the parity even and odd part, respectively. For S=1 there are
two sets of coupled integral equations in two variables for the helicities $\Lambda=0,1$. The
$\phi''$-integration over the potential, Eq.~(\ref{63}),  can be carried out
independently, and thus does not enter the integral equation as separate
variable. As it turns out, both potentials only depend weakly on this angle,
and one typically needs 10 integration points in the case of Bonn-B and 16 in
the case of AV18 to have sufficient accuracy. Typical grids needed for the
$\theta''$-integration consist of 32 to 48 points and for the
$q''$-integration of 48 to 72 points, depending on the desired accuracy.

The Cauchy singularity in Eq.~(\ref{62}) is separated into a principal value
part and a $\delta$-function part, and the principal value singularity is
treated by subtraction. In case of the Bonn-B potential the integration
interval for the $q''$-integration is covered by mapping the Gauss-Legendre
points $u$ from the interval $(0,1)$ via 
\begin{equation}
q= b \tan \left(\frac {\pi}{2}u \right)
\end{equation}
to the interval $(0,\infty)$. Typical values of $b$ are 1000 MeV/c. For the
AV18 potential this type of integration map is less useful due to the difficulty
in handling the numerical Fourier transform for very large values of $q''$,
and the integration interval is cut off at 150~fm$^{-1}$, which is according
to our experience  sufficient. 
The Fourier-Bessel transform of the AV18 potential functions are carried out
using Filon's quadrature formula \cite{asteg}, which proves to be more
accurate in handling the strong oscillatory behavior of the integrands
in Eqs.~(\ref{71})-~(\ref{75}) for
large values of the integration parameter $r$, compared to e.g.
Simpson's rule.

A stringent test for our numerics is a comparison of the NN phase shifts,
which we obtain from  
Eq.~(\ref{88}) together with Eq.~(\ref{86}),  with phase shifts calculated 
with standard partial wave techniques. The result for the Bonn-B phase shifts
is given in Table~I for projectile energies 100 and 300 MeV. The agreement of
both calculations is excellent. In Table~II the equivalent comparison is given between
the phase shifts calculated from the AV18 potential with our 3D formulation and 
a standard partial wave calculation (in this case in coordinate space). The agreement is   
also very good, however not as perfect as in the Bonn-B case. 
The reason for this slight discrepancy is presumably twofold. First, we carry out a numerical
Fourier-Bessel transformation of the different potential functions of AV18. The solution of the t-matrix
integral equation requires an evaluation of the potential function on a grid of the 
size $n_{\phi''} \times n_{\theta''} \times n_{q''} \times n_{q'} \times n_{\theta'}$.
For computational economy we calculate the potential functions on a
fine grid for $\rho=|{\bf q''}-{\bf q}|$ and obtain the points actually needed 
in the calculation via
interpolation. This procedure naturally leads to larger numerical errors compared 
to a direct evaluation
of algebraic expressions as is the case for Bonn-B. The differences can be clearly seen when comparing
Tables~I and II. In both cases a comparable grid for the t-matrices is used.

Next we want to display the angular and momentum dependence of the
half-shell t-matrices $T^{\pi St}_{\Lambda' \Lambda}(q,q_0,\theta)$ as 
generated from Eq.~(\ref{62}). For the case $S=0$ this is done in Fig.~1 for the Bonn-B potential 
for the parity-even (left side) and  for the parity-odd case (right side). 
The parity-even t-matrix exhibits a similar behavior to the
symmetrized t-matrix of the two-boson case studied in Ref.~\cite{trit3d}. It
shows strong forward and backward peaks around the on-shell momentum
$q_0$. Once far away from $q_0$ the momentum dependence is quite
moderate. The parity-odd t-matrix also displays the the strong peaks
around the on-shell momentum, however the forward and backward peaks
have opposite signs. 

For the case $S=1$ there are 
two coupled integral equations as shown in
Eq.~(\ref{62}). Due to the symmetries of the potential, Eq.~(\ref{19}),
it is sufficient to consider only the helicity combinations 
$\Lambda',\Lambda =0,1$. The corresponding half-shell t-matrices,
$T^{\pi 1t}_{\Lambda' \Lambda}(q,q_0,\theta)$  obtained from the Bonn-B
potential are displayed in Figs.~2 through 5. The real and imaginary
parts show a rich structure as function of angle and for momenta smaller
than 1000 MeV/c. For momenta larger than 1000 MeV/c  all amplitudes show 
no or only a very weak angular dependence. We would like to point out that at 300~MeV
projectile energy partial waves have to be summed at least to $J=6$ to obtain a converged
result for calculating e.g. observables. 

A further question of interest is  how strong the difference is between
the half-shell t-matrix
amplitudes derived from the two potentials 
presented here. In Fig.~6 the real and imaginary parts of 
$T^{\pi St}_{\Lambda' \Lambda}(q,q_0,\theta)$
for $S=0$ calculated with AV18 are shown. Of course along the line $q=q_0$ the
corresponding amplitudes derived from the two potentials are identical. However, if
one compares the detailed structures of especially the parity odd amplitudes, one
sees that at large momenta $q$ the Bonn-B amplitude shows more structure than the
one derived from AV18.
For the case S=1 we only want to show a selection of t-matrix amplitudes calculated
from AV18. In general one can say that some amplitudes are very similar to the ones
derived from Bonn-B, others are quite different. The choices displayed in Fig.~7
are being representative. The upper two t-matrices show a considerable difference
to the corresponding ones of Bonn-B, while the lower ones are relatively similar.

Next we turn to the on-the-energy-shell t-matrix amplitudes
as given in Eq.~(\ref{28}) in terms of the \( T^{\pi St}_{\Lambda '\Lambda } \)
quantities.  Here it is interesting to consider 
$|_a\langle \nu_1 \nu_2 m_1' m_2' q_0 \hat{q}|T|\nu_1 \nu_2 m_1 m_2 {\bf q_0}\rangle
_a|^2 \equiv  |\langle m_1' m_2' | T| m_1 m_2 \rangle |^2$, 
where $m=\pm \frac{1}{2}$. If one takes rotational symmetry and parity invariance 
into account, one ends up with six independent amplitudes,
\begin{eqnarray}
\langle ++|T|++\rangle & = & \langle --|T|--\rangle \nonumber \\
\langle ++|T|--\rangle & = & \langle --|T|++\rangle \nonumber \\
\langle +-|T|++\rangle & = & \langle -+|T|++\rangle =
-\langle +-|T|--\rangle =
-\langle -+|T|--\rangle \nonumber \\
\langle ++|T|+-\rangle & = & \langle ++|T|-+\rangle = 
-\langle --|T|+-\rangle = -\langle --|T|-+\rangle \nonumber \\
\langle +-|T|+-\rangle & = & \langle -+|T|-+\rangle \nonumber \\
\langle -+|T|+-\rangle & = & \langle +-|T|-+\rangle 
\end{eqnarray}
The six amplitudes given above are displayed in Figs.~8 to 10 for Bonn-B as function
of energy and c.m. scattering angle $\cos \theta$. Though both potentials are
only meant to be applied below the pion production threshold, we show the on-shell amplitudes
up to 1~GeV. Since we do not work with partial waves, a calculation at higher energies takes the 
same effort as one at very low energies.  The sum over all indices $m_i$ of the on-shell amplitudes
shown gives
the spin averaged differential cross section as indicated in Eq.~(\ref{eq:5.13}).
We would like to remark that these on-shell amplitudes can also be obtained from the
partial wave projected S-matrix elements as indicated in Eq.~(\ref{eq:5.12}). We used
this relation for numerical tests of our formulation.

\section{Summary}
Two nucleon scattering at intermediate energies of a few hundred MeV
requires quite a few angular momentum states in order to achieve convergence
of e.g. scattering observables. This is even more true for the scattering
of three or more nucleons upon each other. An alternative approach to the
conventional one, which is based on angular momentum decomposition, is to
work directly with momentum vectors, specifically with the magnitudes of the
momenta and the angles between them.
We formulated and numerically
illustrated this alternative approach for the case of NN scattering using two realistic
interaction models.
The momentum vectors enter directly into the scattering equation, and the total spin
of the two nucleons is treated in a helicity representation with respect to the relative
momenta of the two nucleons. Using rotational and parity invariance one finds in the
triplet case S=1 a set of two coupled Lippmann-Schwinger equations for each parity and each
initial helicity state. Because of symmetry properties only two of the originally three
initial helicity states need to be considered. 
In the singlet case S=0 there is only one single Lippmann-Schwinger equation 
for each parity.  The Lippmann-Schwinger equations (uncoupled and coupled) are two-dimensional
integral equations in two variables for the half-shell t-matrix and three variables for the
fully-off-shell t-matrix, namely two magnitudes of momenta and one angle.
Though we start with a three-dimensional integration, one angle can be integrated out
independently,
so that we are left with a two-dimensional integral equation.

A formulation without angular momentum decomposition 
is best suited for interaction models which are given in an operator form. 
In this work we considered
the  Bonn-B and the AV18 potentials. The helicity  representation  of the potentials
is given in the six linear operators $\Omega_i$ of Eq.~(\ref{44}), depending on spin and
momenta, which are most suited for our formulation. For completeness we also give the
potentials represented via Wolfenstein operators. We demonstrate the feasibility and accuracy
of our three-dimensional formulation by projecting the on-shell t-matrix elements on angular
momentum states and compare the resulting phase-shifts with those obtained from standard
partial wave projected Lippmann-Schwinger equations. The agreement is very good and
demonstrates the numerical reliability and accuracy of our method.

The complete set of half-shell t-matrix amplitudes at 300~MeV laboratory energy is displayed
for the Bonn-B potential to give an impression of their rich structure in momentum and angle
variables. 
Nevertheless the resulting angular dependence is smoother than the individual contributions
resulting from the higher angular momentum states.

In some representative examples it is shown that those half-shell amplitudes can
differ significantly depending on the potential employed. Of course,
the observable quantities,
the on- shell amplitudes, are equal since they describe the same NN data base.
We display the magnitudes of the physical t-matrix
elements derived from the Bonn-B model  as a function of  scattering angle and energy. 
Though the NN
potential is tuned only up to about 350 MeV our results are shown up to 1 GeV
to demonstrate the ease by which also high energies can be handled in our approach without
partial waves.

We would like to emphasize that the here developed  scheme is algebraically 
very simple to handle provided
potentials are given in an operator form. This is e.g. the case for all interactions
developed within a field theoretic frame work.
In addition, the solution of two-dimensional integral equations does not pose any difficulty
for modern computers.

This work is intended to serve as  starting point towards  treating three-nucleon scattering
without partial waves. In Refs. \cite{trit3d,scatter1} three-nucleon bound and scattering
states  
have already been treated without partial waves, however neglecting spin degrees of
freedom. 
In the present work we show at the two-nucleon level that the inclusion of spin degrees of
freedom is quite readily possible. 
Future work needs to combine the experience with the NN system gained here with the three-body
calculations without partial waves of Refs. \cite{trit3d,scatter1}. This will be an important
step forward since present 3N scattering codes based on partial wave methods reach the limit of
todays supercomputers already at intermediate energies below the pion threshold. Future
applications at higher energies require new techniques.

\vfill

\acknowledgements This work was performed in part under the
auspices of the Deutsche Akademische Austauschdienst under contract No. A/96/32258,
the U.~S.  Department of Energy under contract
No. DE-FG02-93ER40756 with Ohio University, the NATO Collaborative
Research Grant 960892, and the
National Science Foundation under Grant No. INT-9726624.
We thank the Computer Center of the RWTH  Aachen for the use of their
facilities under Grant P039.
 
\newpage    

\appendix

\section{The Transformation Matrix \protect\( A\protect \) }

In this appendix we explicitly give the transformation from the
Wolfenstein operators given in Eqs.~(\ref{33}) to the operators
$\Omega_i$ of Eqs.~(\ref{44}), i.e. the matrix $A_{ij}$ indicated in
Eq.~(\ref{45}). The matrix is explicitly given as
\begin{equation}
A=\left( \begin{array}{cccccc}
1 & 0 & 0 & 0 & 0 & 0\\
0 & \frac{ia}{\gamma } & \frac{-2i}{\gamma a} & \frac{2i}{a} & \frac{2i}{\gamma a} & \frac{-2i}{\gamma a}\\
-1 & 1 & 0 & \frac{2\gamma }{a^{2}} & \frac{-2}{a^{2}} & 0\\
-1 & \frac{-q'qa^{2}}{\gamma c} & \frac{2q'(q'\gamma +q)}{\gamma c} & \frac{2q'q}{c} & \frac{-2q'q}{\gamma c} & \frac{2q(q\gamma +q')}{\gamma c}\\
-1 & \frac{q'qa^{2}}{\gamma b} & \frac{2q'(q'\gamma -q)}{\gamma b} & \frac{-2q'q}{b} & \frac{2q'q}{\gamma b} & \frac{2q(q\gamma -q')}{\gamma b}\\
\frac{-2(q'^{2}-q^{2})}{e} & 0 & \frac{4q'^{2}}{e} & 0 & 0 & \frac{-4q^{2}}{e}
\end{array}\right). 
\end{equation}
The matrix elements $ A_{ij}$ are scalar functions. We calculate
\begin{equation} 
\label{47}
det\, A=\frac{-i2^{9}}{\gamma \sqrt{1-\gamma ^{2}}}\left(
\frac{qq'}{|{\bf q}'+{\bf q}||{\bf q}'-{\bf q}|}\right) ^{3},
\end{equation}
and find the inverse matrix, $A_{ij}^{-1}$ to be
\begin{equation} 
A^{-1}=\left( \begin{array}{cccccc}
1 & 0 & 0 & 0 & 0 & 0\\
\frac{3}{2} & 0 & \frac{1}{2} & \frac{e^2}{8a^2 q'^2 q^2} & \frac{e^2}{8a^2 q'^2 q^2}
& \frac{-(q'^2-q^2)e}{8a^2 q'^2 q^2}\\
\frac{1}{2} & 0 & 0 & \frac{c}{8q'^{2}} & \frac{b}{8q'^{2}} & \frac{e}{8q'^{2}}\\
\frac{\gamma }{2} & \frac{-ia}{4} & 0 & \frac{c}{8q'q} & \frac{-b}{8q'q} & 0\\
\frac{1+\gamma ^{2}}{4} & \frac{-ia\gamma }{4} & \frac{-a^{2}}{4} & \frac{fc}{16} & \frac{fb}{16} & \frac{ge}{16}\\
\frac{1}{2} & 0 & 0 & \frac{c}{8q^{2}} & \frac{b}{8q^{2}} & \frac{-e}{8q^{2}}
\end{array}\right). 
\label{A3}
\end{equation}
For both matrices we introduced the abbreviations
\begin{eqnarray*}
a & = & \sqrt{1-\gamma ^{2}}\\
b & = & q'^{2}+q^{2}-2q'q\gamma \\
c & = & q'^{2}+q^{2}+2q'q\gamma \\
e & = & \sqrt{(q'^{2}+q^{2})^{2}-4q'^{2}q^{2}\gamma ^{2}}\\
f & = & \frac{1}{q'^{2}}+\frac{1}{q^{2}}\\
g & = & \frac{1}{q'^{2}}-\frac{1}{q^{2}}
\end{eqnarray*}

\section{The OBE Potential as Function of Wolfenstein operators}

In terms of the Wolfenstein operators $W_i$ of Eq.~(\ref{34}), the Bonn-B
potential takes the following form:
\begin{eqnarray}
V_{ps}({\bf q}',{\bf q}) & = & \frac{g_{ps}^{2}}{(2\pi
)^{3}4m^{2}}\sqrt{\frac{m}{E'}}\sqrt{\frac{m}{E}}\frac{F^{2}_{ps}[({\bf
q}'-{\bf q})^2]}{({\bf q}'-{\bf q})^{2}+m_{ps}^{2}}\frac{\hat{O}_{ps}}{W'W}
 \\
 &  &  \nonumber \\
V_{s}({\bf q}',{\bf q}) & = & \frac{g_{s}^{2}}{(2\pi
)^{3}4m^{2}}\sqrt{\frac{m}{E'}}\sqrt{\frac{m}{E}}\frac{F^{2}_{s}[({\bf
q}'-{\bf q})^2]}{({\bf q}'-{\bf q})^{2}+m_{s}^{2}}\frac{\hat{O}_{s}}{W'W} \\
 &  &  \nonumber \\
V_{v}({\bf q}',{\bf q}) & = & \frac{1}{(2\pi
)^{3}4m^{2}}\sqrt{\frac{m}{E'}}\sqrt{\frac{m}{E}}\frac{F^{2}_{v}[({\bf
q}'-{\bf q})^2]}{({\bf q}'-{\bf q})^{2}+m_{v}^{2}}
\frac{\left( g_{v}^{2}\hat{O}_{vv}+2g_{v}f_{v}\hat{O}_{vt}+
f_{v}^{2}\hat{O}_{tt}\right) }{W'W} ,
\end{eqnarray}
where $m$ denotes the nucleon mass and $W=\sqrt{m^2+{\bf q}^2}$.
The operators $\hat{O}_{ps,s,v}$ are given as
\begin{eqnarray}
\hat{O}_{ps} & = & -\frac{1}{4}[(W'-W)^{2}(q'^{2}+q^2+2q'q\gamma ){\bf W_4}+
(W'+W)^2 (q'^2+q^2-2q'q\gamma ){\bf W_5}] \nonumber \\
 &  & -(W'^{2}-W^{2})\sqrt{(q'^2+q^2)^{2}-4q'^{2}q^{2}\gamma^{2}} \:{\bf W_6}\\
 &  &  \nonumber \\
\hat{O}_{s} & = & -(W'W-q'q\gamma )^2\: {\bf W_1}-i(W'W-q'q\gamma)
q'q\sqrt{1-\gamma ^{2}} \:{\bf W_2} \nonumber \\
 & & +q'^{2}q^{2}(1-\gamma ^{2})\: {\bf W_3}  \\
 &  &  \nonumber \\
\hat{O}_{vv} & = & (W'^{2}W^{2}+q'^{2}q^{2}\gamma^2+4W'Wq'q\gamma +
W'^{2}q^{2}+W^{2}q'^{2})\: {\bf W_1} \\
 &  & -i(3W'W+q'q\gamma )q'q\sqrt{1-\gamma ^{2}} \:{\bf W_2} \nonumber \\
 &  & -\left[ 1-\gamma ^{2}+\left( W'^{2}q^{2}+W^{2}q'^{2}-2W'Wq'q\gamma \right) 
 \right] {\bf W_3} \nonumber \\
 &  & +\left[ (W'^{2}+W^{2}-2W'W) -\frac{(q'^{2}+q^{2}-2q'q\gamma)}
 {q'^{2}q^{2}(1-\gamma ^{2})}
 ( W'^{2}q^{2}+W^{2}q'^{2}-2W'Wq'q\gamma ) \right] \nonumber  \\
 &  & \times \frac{1}{4} (q'^{2}+q^{2}+2q'q\gamma ) {\bf W_4} \nonumber \\
 &  & +\left[ (W'^2+W^2+2W'W) - \frac{(q'^{2}+q^{2}+2q'q\gamma)}
  {q'^{2}q^{2}(1-\gamma ^{2})}
( W'^{2}q^{2}+W^{2}q'^{2}-2W'Wq'q\gamma \right] \nonumber  \\
 &  & \times \frac{1}{4} (q'^{2}+q^{2}-2q'q\gamma ){\bf W_5}  \nonumber\\
 &  & -\left[ (W'^{2}-W^{2}) -\frac{(q'^2-q^2)}{q'^{2}q^{2}(1-\gamma ^{2})}
( W'^{2}q^{2}+W^{2}q'^{2}-2W'Wq'q\gamma  \right]  \nonumber \\
 &  & \times \frac{1}{4} \sqrt{(q'^{2}+q^{2})^{2}-4q'^{2}q^{2}\gamma ^{2}} {\bf W_6} 
 \nonumber \\
 &  &  \nonumber \\
\hat{O}_{vt} & = & \left[ \frac{(2m-E'-E)}{m}W'^{2}W^{2}+\frac{(W'+W)}{m}q'^{2}q^{2}
\gamma ^{2}+2q'qW'W\gamma \right.  \nonumber \\
 &  & +\left.
\frac{(2m-E'+E)}{2m}W'^{2}q^{2}+\frac{(2m+E'-E)}{2m}W^{2}q'^{2}\right] \;
 {\bf W_1} \\
 &  & -i\left[ \frac{(W'+W)}{m}q'q\gamma +2W'W\right] q'q\sqrt{1-\gamma ^{2}} {\bf W_2}
 \nonumber \\
 &  & -\left\{ \frac{(W'+W)}{m}(1-\gamma ^{2})\right. \nonumber  \\
 &  & +\left. \left[ \frac{(2m-E'+E)}{2m}W'^{2}q^{2}+\frac{(2m+E'-E)}{2m}W^{2}q'^{2}
 -2W'Wq'q\gamma \right] \right\} \; {\bf W_3} \nonumber \\
 &  & +\left\{ \frac{(2m-E'+E)}{2m}W'^{2}+\frac{(2m+E'-E)}{2m}W^{2}-2W'W \right.
  \nonumber  \\
 &  & -\left[ \frac{(2m-E'+E)}{2m}\frac{W'^2}{q'^2}+\frac{(2m+E'-E)}{2m}\frac{W^2}
 {q^{2}}-2W'W\frac{\gamma }{q'q}\right]  \nonumber \\
 &  & \times \left. \frac{(q'^{2}+q^{2}-2q'q\gamma)}{(1-\gamma^2)}\right\}
  \frac{1}{4} (q'^{2}+q^{2}+2q'q\gamma ) \; {\bf W_4} \nonumber \\
 &  & +\left\{ \frac{(2m-E'+E)}{2m}W'^{2}+\frac{(2m+E'-E)}{2m}W^{2}+2W'W \right.
 \nonumber \\
 &  & -\left[ \frac{(2m-E'+E)}{2m}\frac{W'^{2}}{q'^{2}}+\frac{(2m+E'-E)}{2m}
 \frac{W^{2}}{q^{2}}-2W'W\frac{\gamma }{q'q}\right]  \nonumber \\
 &  & \times \left. \frac{(q'^{2}+q^{2}+2q'q\gamma)}{(1-\gamma ^{2})}\right\}
 \frac{1}{4} (q'^{2}+q^{2}-2q'q\gamma) \; {\bf W_5}  \nonumber \\
 &  & -\left\{ \frac{(2m-E'+E)}{2m}W'^{2}-\frac{(2m+E'-E)}{2m}W \right. \nonumber \\
 &  & -\left[ \frac{(2m-E'+E)}{2m}\frac{W'^{2}}{q'^{2}}+\frac{(2m+E'-E)}{2m}
 \frac{W^{2}}{q^{2}}-2W'W\frac{\gamma }{q'q}\right] \nonumber \\
 &  & \times \left. \frac{(q'^{2}-q^{2})}{(1-\gamma ^{2})}\right\}
 \frac{1}{4} \sqrt{(q'^{2}+q^{2})^{2}-4q'^{2}q^{2}\gamma^2} \: {\bf W_6} \nonumber \\
 &  &  \nonumber \\
  \hat{O}_{tt}& = &\left\{ \left[ 5m^{2}-4m(E'+E)+3E'E+q'q\gamma \right] W'^2 W \right. \\
 &  & +\left[ 5m^{2}+4m(E'+E)+3E'E+q'q\gamma \right] q'^{2}q^{2}\gamma^2 \nonumber \\
 &  & +(2m^{2}-4E'E-E'^{2}-E^{2}-2q'q\gamma )W'Wq'q\gamma \nonumber \\
 &  & +\left. \frac{(2m-E'+E)^{2}}{2}q^{2}W'^{2}+\frac{(2m+E'-E)^{2}}{2}q'^{2}W^{2}
 \right\} \frac{1}{2m^2} \: {\bf W_1} \nonumber \\
 &  & -i\left\{ (3m^{2}-E'E-E'^{2}-E^{2}-q'q\gamma )W'W \right. \nonumber \\
 &  & +\left. \left[ 5m^{2}+4m(E'+E)+3E'E+q'q\gamma \right] q'q\gamma \right\} 
  \frac{q'q}{2m^2} \sqrt{1-\gamma^2} \; {\bf W_2}  \nonumber \\
 &  & -\left\{ [10m^{2}+8m(E'+E)+6E'E+2q'q\gamma ](1-\gamma ^{2}) \right.
\nonumber \\
 &  & +\left[(2m-E'+E)^{2}W'^2 q^2+(2m+E'-E)^{2}W^{2}q'^{2}\right. \nonumber \\
 &  & -\left. \left. 2[4m^{2}-(E'-E)^{2}]W'Wq'q\gamma \right] \right\} 
  \frac{W_{3}}{4m^{2}} \nonumber \\
 &  & +\left\{ (2m-E'+E)^{2}W'^{2}+(2m+E'-E)^{2} W^2 \right. \nonumber \\
 &  & -2[4m^{2}-(E'-E)^{2}]W'W \nonumber \\
 &  & -\left[ (2m-E'+E)^{2}\frac{W'^{2}}{q'^{2}}+(2m+E'-E)^{2}\frac{W^{2}}{q^{2}}\right.
 \nonumber  \\
 &  & -\left. 2[4m^{2}-(E'-E)^{2}]W'W\frac{\gamma }{q'q} \right] \nonumber \\
 &  & \times \left. \frac{(q'^{2}+q^{2}-2q'q\gamma)}{(1-\gamma ^{2})}\right\}
  \frac {(q'^{2}+q^{2}+2q'q\gamma)}{16 m^2} \; {\bf W_4} \nonumber \\ 
 &  & +\left\{ (2m-E'+E)^{2}W'^{2}+(2m+E'-E)^{2}W^2 \right. \nonumber \\
 &  & +2[4m^{2}-(E'-E)^{2}]W'W \nonumber \\
 &  & -\left[ (2m-E'+E)^{2}\frac{W'^{2}}{q'^{2}}+(2m+E'-E)^{2}\frac{W^{2}}{q^{2}}\right.
 \nonumber \\
 &  & -\left. 2[4m^{2}-(E'-E)^{2}]W'W\frac{\gamma }{q'q} \right] \nonumber \\
 &  & \times \left. (q'^{2}+q^{2}+2q'q\gamma )\frac{1}{(1-\gamma ^{2})}\right\}
 \frac{(q'^{2}+q^{2}-2q'q\gamma)}{16m^2} \; {\bf W_5} \nonumber \\
 &  & -\left\{ (2m-E'+E)^{2}W'^{2}-(2m+E'-E)^{2}W^2 \right. \nonumber \\
 &  & -\left[ (2m-E'+E)^{2}\frac{W'^{2}}{q'^{2}}+(2m+E'-E)^{2}\frac{W^{2}}{q^{2}}\right.
 \nonumber \\
 &  & -\left. 2[4m^{2}-(E'-E)^{2}]W'W\frac{\gamma }{q'q} \right] \nonumber \\
 &  & \times \left. \frac{(q'^{2}-q^{2})}{(1-\gamma ^{2})}\right\} 
 \frac{\sqrt{(q'^{2}+q^{2})^{2}-4q'^{2}q^{2}\gamma^2}}{16m^2} \; {\bf W_6}
\nonumber
\end{eqnarray}
For the actual calculations the above given terms are re-expressed in
terms of the operators $\Omega_i$ via $\Omega_i=\sum_j A_{ij}^{-1} W_j$ 
and inserted into the L.S. equation Eq.~(\ref{62}). These reformulations are
straightforward, and can e.g. be carried out via symbolic manipulation packages.

\section{The Argonne AV18 Potential as Function of Wolfenstein Operators}

In terms of the Wolfenstein operators $W_i$ of Eq.~({\ref{34}), the 
AV18 potential takes the following form:
\begin{eqnarray}
v^c_{ST}({\bf q}',{\bf q}) & = & {\bf W_1} \frac{1}{2\pi^2} \; \int_0^{\infty} dr \ r^2
    j_0(\rho r) v^c_{ST}(r) \\
& &  \nonumber \\
v^t_{ST}({\bf q}',{\bf q}) & = & \Biggl[ {\bf W_3} -\frac{1}{4}\frac{({\bf q'}-{\bf q})^2
 ({\bf q'}+{\bf q})^2} {q'^2q^2(\gamma^2-1)} \ {\bf W_4} 
- \frac{1}{4}\frac{(q'^4+8q'^2q^2\gamma^2-10q'^2q^2+4q^2)}
  {q'^2q^2(\gamma^2-1)} \ {\bf W_5}  \nonumber \\
&+& \frac{1}{4} \frac{\sqrt{(q'^2+q^2)^2-4q'^2q^2\gamma^2}(q'^2-q^2)}
   {q'^2q^2(\gamma^2-1)} \ {\bf W_6} \Biggr]
  \frac{1}{2\pi^2}\frac{1}{\rho^2} 
  \; \int_0^{\infty} dr \ r^2 j_2(\rho r) v^t_{ST}(r) \\
& &  \nonumber \\
v^{ls}_{ST}({\bf q}',{\bf q}) & = & \frac{i}{2} q'q \sqrt{1-\gamma^2} \ {\bf W_2}
  \frac{1}{2\pi^2} \frac{1}{\rho} \; \int_0^{\infty} dr \ r^3 j_1(\rho r) v^{ls}_{ST}(r) \\
& &  \nonumber \\
v^{l2}_{ST}({\bf q}',{\bf q}) & = & q'q\gamma {\bf W_1} \frac{1}{2\pi^2} \frac{2}{\rho}
 \; \int_0^{\infty} dr \ r^3 j_1(\rho r) v^{l2}_{ST}(r)  \nonumber \\
 & & - (q'^2q^2(1-\gamma^2)) {\bf W_1} \frac{1}{2\pi^2} \frac{1}{\rho^2}
   \; \int_0^{\infty} dr \ r^4 j_2(\rho r) v^{l2}_{ST}(r)   \\
& &  \nonumber \\
v^{ls2}_{ST}({\bf q}',{\bf q}) & = & \Biggl[ q'q\gamma {\bf W_1} -\frac{i}{4}q'q\sqrt{1-\gamma^2}
 {\bf W_2} + \frac{1}{2} q'q\gamma {\bf W_3} \nonumber \\
&-& \frac{1}{8} \frac{({\bf q'}+{\bf q})^2(q'-\gamma q)(q'\gamma-q)}
  {q'q(\gamma^2-1)} {\bf W_4} \nonumber \\
&-& \frac{1}{8} \frac{({\bf q'}-{\bf q})^2(q'+q\gamma)(q+q'\gamma)}{q'q(\gamma^2-1)} {\bf W_5}
+ \frac{1}{8} \frac{\gamma (q'^2-q^2) \sqrt{(q'^2+q^2)^2 -4q'^2q^2\gamma^2}} 
 {q'q(\gamma^2-1)} {\bf W_6} \Biggl]
 \nonumber \\
& &  \frac{1}{2\pi^2} \frac{1}{\rho} \; \int_0^{\infty} dr \ r^3 j_1(\rho r)v^{ls2}_{ST}(r)
 \nonumber \\
&+& \frac{1}{2}q'^2q^2(\gamma^2-1) ( {\bf W_1} + {\bf W_3}) \frac{1}{2\pi^2}
\frac{1}{\rho^2} \; \int_0^{\infty} dr \ r^4 j_2(\rho r) v^{ls2}_{ST}(r)
\end{eqnarray}
In our actual calculation  the potential enters  described by the 
set of operators
$\Omega_i$ of Eq.~({\ref{44}). In order to display the relative simple form the 
potential takes in these operators, we also want to explicitly show them.
\begin{eqnarray}
v^c_{ST}({\bf q}',{\bf q}) & = &  {\bf \Omega_1} \frac{1}{2\pi^2} \; \int_0^{\infty} dr \ r^2
 j_0(\rho r) v^c_{ST}(r) \\
v^t_{ST}({\bf q}',{\bf q}) & = & 
  \Biggl[6q'q {\bf \Omega_4} + 2q'^2 ({\bf \Omega_2} -3{\bf \Omega_3}) +2q^2
   ({\bf \Omega_2} -3{\bf \Omega_6}) -q'q\gamma {\bf \Omega_2} \nonumber \\
& & -3q'q\frac{1}{\gamma}({\bf \Omega_2}-2{\bf \Omega_3}-2{\bf \Omega_6}+2{\bf \Omega_5)}) \Biggr]
        \frac{1}{2\pi^2} \frac{1}{\rho^2} \; 
 \int_0^{\infty} dr \ r^2 j_2(\rho r) v^t_{ST}(r)  \\
v^{ls}_{ST}({\bf q}',{\bf q}) & = & -\frac{1}{2} q'q \Biggl[ 2 {\bf \Omega_4} 
   -\gamma {\bf \Omega_2} +\frac{1}{\gamma}({\bf \Omega_2} -2{\bf \Omega_3}-2{\bf \Omega_6}
  +2{\bf \Omega_5})\Biggr]  \nonumber \\
 & & \;\;\;\frac{1}{2\pi^2} \frac{1}{\rho} \; \int_0^{\infty} dr \ r^3 j_1(\rho r)
v^{ls}_{ST}(r) \\
v^{l2}_{ST}({\bf q}',{\bf q}) & = & 
  q'q\gamma {\bf \Omega_1} \frac{1}{2\pi^2} \frac{2}{\rho}
 \; \int_0^{\infty} dr \ r^3 j_1(\rho r) v^{l2}_{ST}(r)  \nonumber \\
 & & - {\bf \Omega_1} \left[ q'^2q^2(1-\gamma^2)\right] 
    \frac{1}{2\pi^2} \frac{1}{\rho^2}
   \; \int_0^{\infty} dr \ r^4 j_2(\rho r) v^{l2}_{ST}(r)  \\
v^{ls2}_{ST}({\bf q}',{\bf q}) & = & \frac{1}{2} q'q \left( \gamma {\bf \Omega_2}
+\frac{1}{\gamma} ({\bf \Omega_2} +2 {\bf \Omega_5} -2 {\bf \Omega_3}-2{\bf \Omega_6})\right) 
   \frac{1}{2\pi^2} \frac{1}{\rho} \; \int_0^{\infty} dr \ r^3 j_1(\rho r) 
v^{ls2}_{ST}(r)
 \nonumber \\
 & & -\frac{1}{2}q'^2 q^2 \left[(1-\gamma^2){\bf \Omega_2} +2\gamma 
  {\bf \Omega_4} -2{\bf \Omega_5} \right] \frac{1}{2\pi^2} \frac{1}{\rho^2}
 \; \int_0^{\infty} dr \ r^4 j_2(\rho r) v^{ls2}_{ST}(r)  
\end{eqnarray}




\begin{table}
\caption{Comparison of the nucleon-nucleon phase shifts calculated from
our three-dimensional formulation (helicitiy)  with a standard 
partial wave calculation (carried out in momentum space)
 for the Bonn-B potential at 100 and 300
MeV laboratory energy.}
\begin{tabular}{|c||r|r||r|r|}
\hline
 ~~~~  &
\multicolumn{2}{c||}{\( E_{lab}=100\, MeV \)}&
\multicolumn{2}{c|}{\( E_{lab}=300\, MeV \)}\\
\hline 
 $^{2S+1}$ L$ _J$ &
3D helicity&
partial wave&
3D helicity.&
partial wave\\
\hline 
\( ^{1}S_{0} \)&
25.1928&
25.1929&
-8.1755&
-8.1756\\
\( ^{3}P_{0} \)&
9.8046 &
9.8046 &
-11.4799 &
-11.4799 \\
\hline 
\( ^{1}P_{1} \)&
-16.3131 &
-16.3451 &
-28.6946 &
-28.8747 \\
\( ^{3}P_{1} \)&
-13.4677 &
-13.4677 &
-26.3800 &
-26.3800 \\
\( ^{3}S_{1} \)&
41.9858 &
41.9870 &
4.0667 &
4.0676 \\
\( ^{3}D_{1} \)&
-12.9847 &
-12.9846 &
-23.7182 &
-23.7181 \\
\( \varepsilon_{1} \)&
-2.2360 &
-2.2357 &
-4.0268 &
-4.0265 \\
\hline 
\( ^{1}D_{2} \)&
3.3411 &
3.3411 &
7.4888 &
7.4888 \\
\( ^{3}D_{2} \)&
17.6710 &
17.6710 &
25.3616 &
25.3617 \\
\( ^{3}P_{2} \)&
11.7356 &
11.7356 &
17.3981 &
17.3981 \\
\( ^{3}F_{2} \)&
0.7705 &
0.7705 &
0.5236 &
0.5238 \\
\( \varepsilon_{2} \)&
2.8402 &
2.8402 &
2.0166 &
2.0166 \\
\hline 
\( ^{1}F_{3} \)&
-2.4397 &
-2.4397 &
-5.5865 &
-5.5865 \\
\( ^{3}F_{3} \)&
-1.6484 &
-1.6484 &
-4.0097 &
-4.0097 \\
\( ^{3}D_{3} \)&
0.4203 &
0.4855 &
2.5719 &
2.5720 \\
\( ^{3}G_{3} \)&
-1.0105 &
-1.0105 &
-4.4051 &
-4.4051 \\
\( \varepsilon_{3} \)&
-3.6604 &
-3.6604 &
-7.2233 &
-7.2233 \\
\hline 
\( ^{1}G_{4} \)&
0.4092 &
0.4092 &
1.3556 &
1.3556 \\
\( ^{3}G_{4} \)&
2.2624 &
2.2624 &
7.3000 &
7.3000 \\
\( ^{3}F_{4} \)&
0.4203 &
0.4203 &
2.4491 &
2.4491 \\
\( ^{3}H_{4} \)&
0.1082 &
0.1082 &
0.5077 &
0.5077 \\
\( \varepsilon_{4} \)&
0.5575 &
0.5575 &
1.5509 &
1.5509 \\
\hline 
\end{tabular}
\end{table}

\begin{table}
\caption{Comparison of the nucleon-nucleon phase shifts calculated from
our three-dimensional formulation (helicity)  with a standard 
partial wave calculation (carried out in coordinate space \protect\cite{bwpers})
 for the AV18 potential at 100 and 300
MeV laboratory energy.}
\begin{tabular}{|c||r|r||r|r|}
\hline
 ~~~~  &
\multicolumn{2}{c||}{\( E_{lab}=100\, MeV \)}&
\multicolumn{2}{c|}{\( E_{lab}=300\, MeV \)}\\
\hline
 $^{2S+1}$ L$ _J$ &
3D helicity&
partial wave&
3D helicity.&
partial wave\\
\hline
\( ^{1}S_{0} \)&
25.99&
 25.94&
-4.62&
 -4.60\\
\( ^{3}P_{0} \)&
8.69&
 8.69&
-11.05&
 -11.06\\
\hline
\( ^{1}P_{1} \)&
-14.19&
 -14.20&
-26.18&
 -26.28\\
\( ^{3}P_{1} \)&
-13.06&
 -13.07&
-28.38&
 -28.49\\
\( ^{3}S_{1} \)&
43.69&
 43.56&
8.15&
 8.16\\
\( ^{3}D_{1} \)&
-12.08&
 -12.09&
-24.80&
 -24.90\\
\( \varepsilon_{1} \)&
-2.49&
 -2.49&
-4.38&
 -4.39\\
\hline 
\( ^{1}D_{2} \)&
3.81&
 3.81&
9.45&
 9.44\\
\( ^{3}D_{2} \)&
17.14&
 17.10&
25.11&
 25.02\\
\( ^{3}P_{2} \)&
11.02&
 11.00&
16.96&
 16.91\\
\( ^{3}F_{2} \)&
0.67&
 0.67&
0.77&
 0.76\\
\( \varepsilon_{2} \)&
2.70&
 2.70&
2.21&
 2.21\\
\hline 
\( ^{1}F_{3} \)&
-2.23&
 -2.23&
-4.87&
 -4.88\\
\( ^{3}F_{3} \)&
-1.35&
 -1.35&
-2.51&
 -2.51\\
\( ^{3}D_{3} \)&
1.61&
 1.61&
5.22&
 5.21\\
\( ^{3}G_{3} \)&
-0.93&
 -0.93&
-4.19&
 -4.20\\
\( \varepsilon_{3} \)&
-3.50&
 -3.50&
-7.17&
 -7.16\\
\hline
\( ^{1}G_{4} \)&
0.40&
 0.40&
1.42&
 1.42\\
\( ^{3}G_{4} \)&
2.22&
 2.22&
7.35&
 7.34\\
\( ^{3}F_{4} \)&
0.45&
 0.45&
2.75&
 2.74\\
\( ^{3}H_{4} \)&
0.07&
 0.07&
0.31&
 0.31\\
\( \varepsilon_{4} \)&
0.51&
 0.51&
1.54&
 1.54 \\
\hline 
\end{tabular}
\end{table}

 
\noindent
\begin{figure}
\caption{The half-shell t-matrix $T^{\pi St}_{\Lambda' \Lambda}(q,q_0,\theta)$ 
for $q_0$=375~MeV/c, 
corresponding to $E_{lab}$=300~MeV, calculated from the Bonn-B potential.
The left side 
displays the real and imaginary parts of $T^{101}_{00}(q,q_0,\theta)$
as function of $q$ and $x=\cos \theta$.
The right side shows $T^{-100}_{00}(q,q_0,\theta)$.
The units of $T^{\pi 0t}_{00}(q,q_0,\theta)$ are $10^{-7}$~MeV$^{-2}$.\label{fig1}}
\end{figure}

\noindent
\begin{figure}
\caption{The real part of the  parity-even half-shell
t-matrix  $T^{110}_{\Lambda'\Lambda}(q,q_0,\theta)$ for
$q_0$=375~MeV/c. The helicities $\Lambda',\Lambda=0,1$
are kept as subscripts of T.
All t-matrices are calculated
using the Bonn-B potential and are given in units $10^{-7}$~ MeV$^{-2}$. \label{fig2}}
\end{figure}

\noindent
\begin{figure}
\caption{Same as Fig.~2, but for the imaginary part of
$T^{110}_{\Lambda'\Lambda}(q,q_0,\theta)$. \label{fig3}}
\end{figure}

\noindent
\begin{figure}
\caption{The real part of the half-shell parity-odd  half-shell
t-matrix  $T^{-111}_{\Lambda'\Lambda}(q,q_0,\theta)$ for
$q_0$=375~MeV/c  calculated from the Bonn-B potential.
The description is the same as in Fig.~2.
\label{fig4}}
\end{figure}

\noindent
\begin{figure}
\caption{Same as Fig.~4, but for the imaginary part of
$T^{-111}_{\Lambda'\Lambda}(q,q_0,\theta)$. \label{fig5}}
\end{figure}

\noindent
\begin{figure}
\caption{The half-shell t-matrix $T^{\pi 0t}_{\Lambda' \Lambda}(q,q_0,\theta)$ 
for $q_0$=375~MeV/c 
calculated from the AV18 potential.
The left side 
displays the real and imaginary parts of $T^{101}_{00}(q,q_0,\theta)$
as function of $q$ and $x=\cos \theta$.
The right side shows $T^{-100}_{00}(q,q_0,\theta)$.
The units of $T^{\pi 0t}_{00}(q',q_0,\theta)$ are $10^{-7}$~MeV$^{-2}$.
 \label{fig6}}
\end{figure}

\noindent
\begin{figure}
\caption{Selected half-shell t-matrices $T^{\pi 1t}_{\Lambda' \Lambda}(q,q_0,\theta)$ 
for $q_0$=375~MeV/c
calculated from the AV18 potential.
The helicities $\Lambda',\Lambda=0,1$ are kept as subscripts of T,
which is given in  units of
$10^{-7}$~MeV$^{-2}$. \label{fig7}}
\end{figure}

\noindent
\begin{figure}
\caption{The absolute value squared of the physical on-shell T-matrix elements
$|\langle m_1' m_2' | T| m_1 m_2 \rangle |^2$ in 
units $10^{-14}$~MeV$^{-4}$. The values of $m=\pm \frac{1}{2}$ are
abbreviated as $\pm$. \label{fig8}} 
\end{figure}

\noindent
\begin{figure}
\caption{Same as Fig.~8, but for different $m$ combinations. \label{fig9}}
\end{figure}

\noindent
\begin{figure}
\caption{ Same as Fig.~8, but for different $m$ combinations.
\label{fig10}}
\end{figure}

\end{document}